\documentclass[article]{aa}
\usepackage{graphicx}

\usepackage{bm}
\usepackage{natbib}
\usepackage{txfonts}
\usepackage{url}
\usepackage{sidecap}

\newcommand{\CaII }{Ca\,II~}
\newcommand{\FeI }{Fe\,I~}

\begin{document}
\bibliographystyle{aa}
\input epsf

\title{The solar chromosphere at high resolution with IBIS}
\subtitle{ II.  Acoustic shocks in the quiet internetwork and the role of magnetic fields}

\author{A. Vecchio\inst{1,2}
\and G. Cauzzi\inst{1,3}
\and K. P.  Reardon\inst{1,3}}

\institute{INAF - Osservatorio Astrofisico di Arcetri, 50125 Firenze, Italy
\and Dipartimento di Fisica, Universit\`a della Calabria, 87036 Rende (CS), Italy
\and National Solar Observatory, P.O. Box 62, Sunspot NM, USA}

\date{\today}

\abstract{The exact nature of the quiet solar chromosphere, and especially its temporal 
variations, are still subject of intense debate. One of the contentious issues is the possible role 
of magnetic field in structuring the quieter solar regions.}
{We characterize the dynamics of the quiet inter-network chromosphere by studying 
the occurrence of acoustic shocks and their  relation with the concomitant photospheric 
structure and dynamics, including small scale magnetic structures.} 
{We analyze a comprehensive data set that includes high resolution chromospheric (\CaII 854.2 
nm) and photospheric (\FeI 709.0 nm) spectra obtained with the IBIS imaging spectrometer in 
two quiet-Sun regions. This is complemented by high-resolution sequences of MDI 
magnetograms of the same targets. From the chromospheric spectra we identify the spatio-
temporal occurrence of the acoustic shocks. We compare it with the photospheric dynamics by 
means of both Fourier and wavelet analysis, and study the influence of magnetic structures on 
the phenomenon.} 
{Mid-chromospheric shocks occur within the general chromospheric dynamics pattern of 
acoustic waves propagating from the photosphere. In particular, they appear as a response to 
underlying powerful photospheric motions at periodicities nearing the acoustic cut-off, 
consistent with 1-D hydrodynamical modeling. However, their spatial distribution within the 
supergranular cells is highly dependent on the local magnetic topology, both at the network and 
internetwork scale. We find that large portions of the internetwork regions undergo very few 
shocks, as ``shadowed'' by the horizontal component of the magnetic field. The 
latter is betrayed by the presence of chromospheric fibrils, observed in the core of the \CaII line 
as slanted structures with distinct dynamical properties. The shadow mechanism 
appears to operate also on the very small scales of inter-network magnetic elements, and 
provides for a very pervasive influence of the magnetic field even in the quietest region 
analyzed. }
{The magnetic field might play a larger role in structuring the quiet solar chromosphere than 
normally assumed. The presence of fibrils highlights a clear disconnection between the 
photospheric dynamics and the response of the geometrically overlaying chromosphere.  As 
these results hold for  a mid-chromospheric indicator such as the \CaII 854.2 line, it is expected 
that diagnostics formed in higher layers, such as UV lines and continua, will be affected to a 
larger extent by the presence of magnetic fields, even in quiet regions. This is of relevance for 
the  chromospheric models that make use of such diagnostics.}

\keywords{Sun: chromosphere --- Sun: magnetic fields --- Sun: oscillations} 
\maketitle
\titlerunning{}
\authorrunning{Vecchio et al.}

\section{Introduction}

Observations have long shown that the outer solar atmosphere is not in radiative
equilibrium, with the net radiative loss of the quiet chromosphere estimated at between 
$\approx$ 4 kWm$^{-2}$ \citep{avrett81} and 14 kWm$^{-2}$ \citep{anderson_89}. The 
identification of the source and exact mechanism of deposition of the energy necessary to 
maintain a stationary situation is often referred to as the problem of chromospheric heating. It 
has been a vexing problem in solar physics for decades.

In stellar atmospheres, strong density stratification makes it relatively easy for propagating 
acoustic waves, excited by the turbulent convection, to develop into shocks. Indeed this mechanism 
was proposed early on as a plausible means to provide the necessary energy input to the 
solar chromosphere \citep{biermann_48,schwarzschild_48}. 
Much theoretical and numerical work has since been conducted to assess the 
viability of the acoustic heating as the basic heating mechanism for nonmagnetic 
chromospheres of slowly rotating stars \citep[e.g.][]{narain_96,ulmrew_03}.

A landmark work on this topic was the 1-D radiative-hydrodynamical modeling of 
\citet{carlsson_95,carlsson_97,carlsson_02}, who derived the chromospheric response to 
acoustic waves
in the absence of magnetic fields. Using an observed photospheric piston, they could  
reproduce the temporal evolution of the chromospheric \CaII H spectral line over long intervals  
with remarkable accuracy. In particular, they reproduced the occurrence of the H$_{2V}$  (K$_{2V}$) grains observed in the spectral profile of the resonance H and K lines
of of Ca\,II.
The grains consist of quasi-periodic  brightenings in the violet wings
of the lines, with a $\sim$150--200 s repetition rate, and are associated with strong redshifts of the line 
core position \citep[see][hereafter RU91, and Section \ref{s_evid_shocks}]{rutten_91}. \citet
{carlsson_97}, hereinafter CS97, showed how the grains are due to weak acoustic shocks 
developing at heights of about 1 Mm in the non magnetic atmosphere, driven primarily by 
photospheric acoustic waves with periodicities close to the cutoff value of $\sim$5.5 mHz. 
A controversial by-product of their model is that the
intermittent acoustic shocks greatly enhance both the emissivity and the temperature but 
only locally, without causing a net outward increase in the {\it average} chromospheric 
temperature \citep{carlsson_95}.  If the CS97 shocks correctly represent the acoustic heating 
mechanism, the non-magnetic chromosphere would be essentially cold. This is contrary to the 
common view, derived from semi-empirical static models, of a chromosphere characterized  
everywhere by a steady temperature rise \citep
{avrett81,fontenla07,avrett_08}. The failure of the CS97 model to reproduce 
spatially and temporally resolved quiet Sun UV observations obtained with SUMER in 
chromospheric lines \citep{carlssonetal_97}
 has further sharpened the controversy.

A strong debate has subsequently ensued about the possible shortcomings of the CS97 model and the 
general viability of the acoustic heating mechanism in shaping the quiet solar chromosphere. 
Much discussion has been devoted to the role of  high frequency acoustic waves \citep[among 
many others,][]{fossum_05,wedemeyeretal_07,cuntz_07,carlssonetal_07}, 
the need to address the problem in 3-D \citep[e.g.][]{wedemeyer_04,ulmschneider_05}, 
and the role of magnetic fields in the quiet Sun \citep[e.g.][]
{judge_03,mcintosh_03,carlsson_07,wedemeyeretal_07}. 
At the bottom of the debate lies the fundamental question of whether the  internetwork 
chromosphere is wholly dynamic in nature \citep{carlsson_07,martinez_07,wedemeyer_07}
 or whether the dynamic variations represent only minor perturbations on a semi-static state of the kind 
described in semi-empirical models \citep{kalkofen_99,avrett_08}.

\begin{figure*}
\centerline{
\includegraphics[scale=.6]{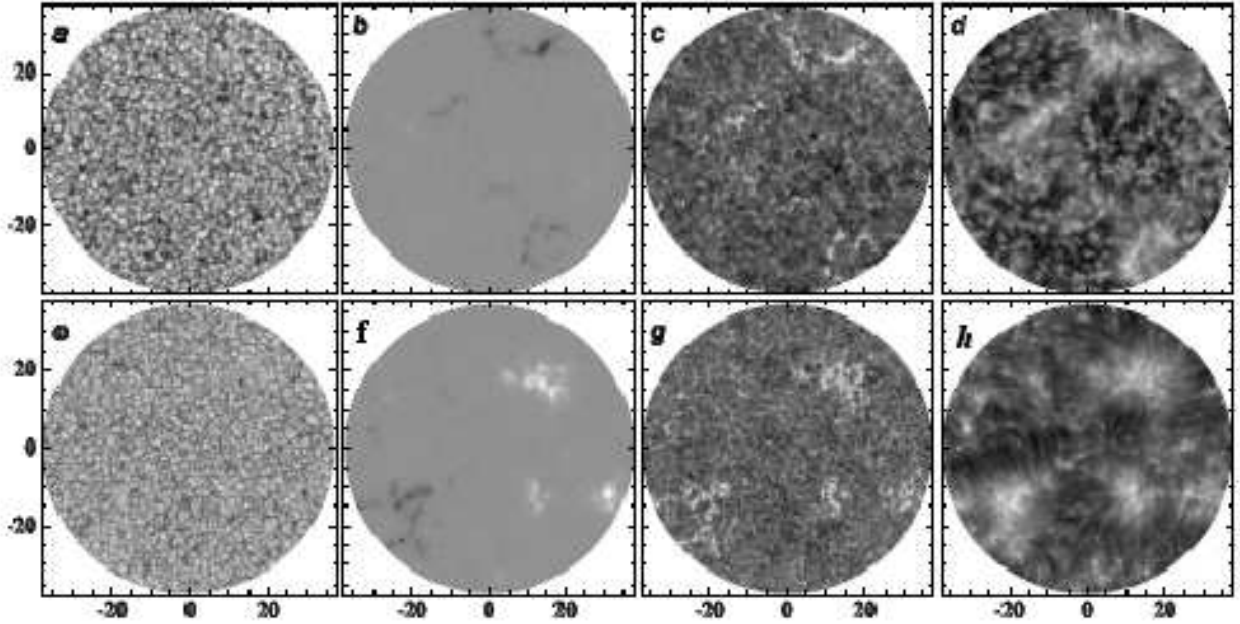}
}
\caption{Top row: data set 1; bottom row: data set 2. The images refer to data acquired around 
the middle of the observing sequences. Axes are given in arcsec.
Panels {\it a, e}: Broadband continuum at 710 nm. The image in data set 2 has been speckle 
reconstructed with the technique of \cite{woeger_06}.
Panels {\it b, f}: co-temporal HR MDI maps, saturated at $\pm$ 500 G. Data set 2 attains to an 
enhanced, bi-polar network region. 
Panels {\it c, g}: Intensity in the red wing of the CaII 854.2 nm at about 0.1 nm from line core. 
Bright points clearly correspond to small scale magnetic features.
Panels {\it d, h}: Line core intensity of CaII 854.2 nm. Note the large extension of fibrils originating in the 
magnetic elements.
}
\label{fig_fov}
\end{figure*}

From the observational point of view, 
gaining further insight into this issue requires new data that can simultaneously achieve 
high spatial and 
temporal resolution, and spectral information with different diagnostics, over extended fields of view (FOV).
Most of the debate to this point has been based on (older)
spectrographic observations obtained in the \CaII H and K lines, and on UV spectral 
data obtained by Skylab or, more recently, by SUMER on board SOHO \citep{wilhelm_95}. 
Spectrographic data suffer however from a limited FOV in fixed slit observations, or from 
a low cadence when scanning an extended region.  In the UV,  
the limited flux and instrumental response further lower the achievable spatial and temporal 
resolution, an important limitation if the dynamic picture just described for the 
chromosphere is the correct one \citep{wedemeyer_07}.  Historically,
imaging with broad-band filters has been widely used, both in the UV and visible range.
Such data, however, often mix signals arising from vastly different regions of the solar atmosphere, as 
well as miss signatures of strong velocities that might shift the lines outside the filter band.
For a 
general review on the issues of chromospheric observations we refer to \citet{rutten_07}, only 
adding here that recently the POLIS spectrograph \citep{beck_05}
has become operational, producing new high quality \CaII H spectra that have been 
employed in studies of the quiet chromosphere \citep{rezaei_07,beck_08,rezaei_08}.

In this paper  we use an entirely different type of instrument, the Interferometric BIdimensional 
Spectrometer \citep[IBIS,][]{cavallini_06}, 
to derive novel results about the dynamics 
of the quiet solar chromosphere and in particular the presence 
(or absence) of acoustic shocks. IBIS is an imaging spectrometer installed at the 
Dunn Solar Telescope (DST) of the US National Solar Observatory, and provides observations 
that combine high spatial and temporal resolution over an extended two-dimensional FOV, with the full 
spectral information in both photospheric and chromospheric lines. In this sense, it is an ideal 
instrument to overcome many of the observational shortcomings outlined above.  We analyze 
the behavior of the chromospheric \CaII 854.2 nm line which, as shown in \citet{cauzzi_08} 
(hereafter Paper I), is one of the most promising diagnostics for high resolution chromospheric 
studies. Equally important, the availability of spectral information over a 2-D FOV is crucial to 
understanding the very dynamic and non-local chromosphere. We further combine the 
 \CaII 854.2 observations with simultaneous and cospatial photospheric data also obtained with 
IBIS and magnetic data from  MDI \citep{scherrer_95}. 
This allows us to address the relationships between the chromospheric and photospheric 
dynamics, and the influence of magnetic field.
We analyze two completely analogous data sets obtained on separate 
days, both times in quiet regions near disk center, and in so doing uncover the fundamental 
role played by the local magnetic topology.
 
The paper is organized as follows. In Section \ref{s_obs} we describe the observations. 
Sections \ref{s_evid_shocks} and \ref{s_findshocks} provide evidence that mid-chromospheric 
acoustic shocks are clearly observed in the \CaII 854.2 nm
line, and outline the methods we use to identify them. In Sect. \ref{s_properties} we report on 
the derived shock properties, both  in relation to earlier findings obtained primarily  with \CaII K 
observations, and shock15.gc.as new results regarding their spatial distribution. Sections \ref{s_mag} and 
\ref{s_phot}  address the occurrence of the shocks in relation to the magnetic field structure 
and evolution, and the photospheric dynamics. Finally, in Sections \ref{s_discussion} and \ref
{s_concl} we discuss our findings and provide conclusions.

\begin{figure}
\centerline{
\includegraphics[scale=0.75]{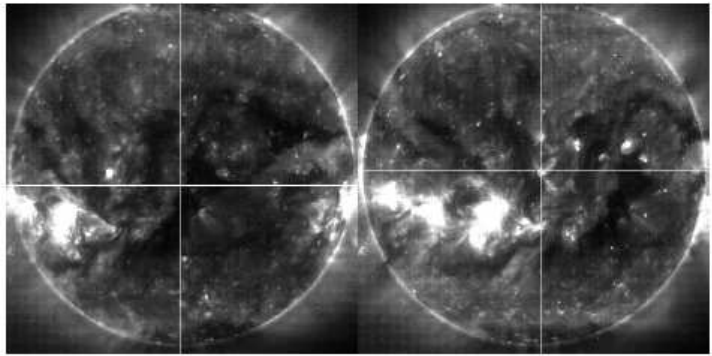}
}
\caption{Full disk EIT images acquired in Fe\,XII 19.5 nm on 2004 May 31 (left) and 2004 June 
02 (right). The cross-hairs indicate the center of the IBIS FOV for each day. The IBIS FOV 
covers roughly the small coronal bright point intersected by the cross-hair on the right panel.}
\label{fig_foveit}
\end{figure}

\section{Observations} \label{s_obs}
The IBIS design and general issues on data reduction have been described in earlier papers 
\citep{cavallini_06,reardon_cavallini_08,janssen_06}, so we mention here only the characteristics 
most relevant to the present work.

IBIS is based on two air spaced Fabry Perot Interferometers in a classical mount, and acquires  
quasi-monochromatic images in the range 560-860 nm (FWHM = 2--4.5 pm). Rapid 
sequential tuning to different
wavelengths allows the spectral sampling of a number of selected lines,
providing spectral information over a circular, 80$"$ diameter
FOV. Typically, a spectral line is sampled in 10--30 spectral points, within a total acquisition 
time of 3--10 s. Data can be obtained at the full resolution of  0.083"/\,pixel, or binned to 
increase the photon flux and reduce the CCD readout time.
Coupled with the  high-order adaptive optics 
system of on the DST \citep{rimmele_04}, IBIS images often attain spatial resolution 
close to the diffraction
limit of the 76-cm telescope. 

\begin{figure*}
\includegraphics[scale=0.6]{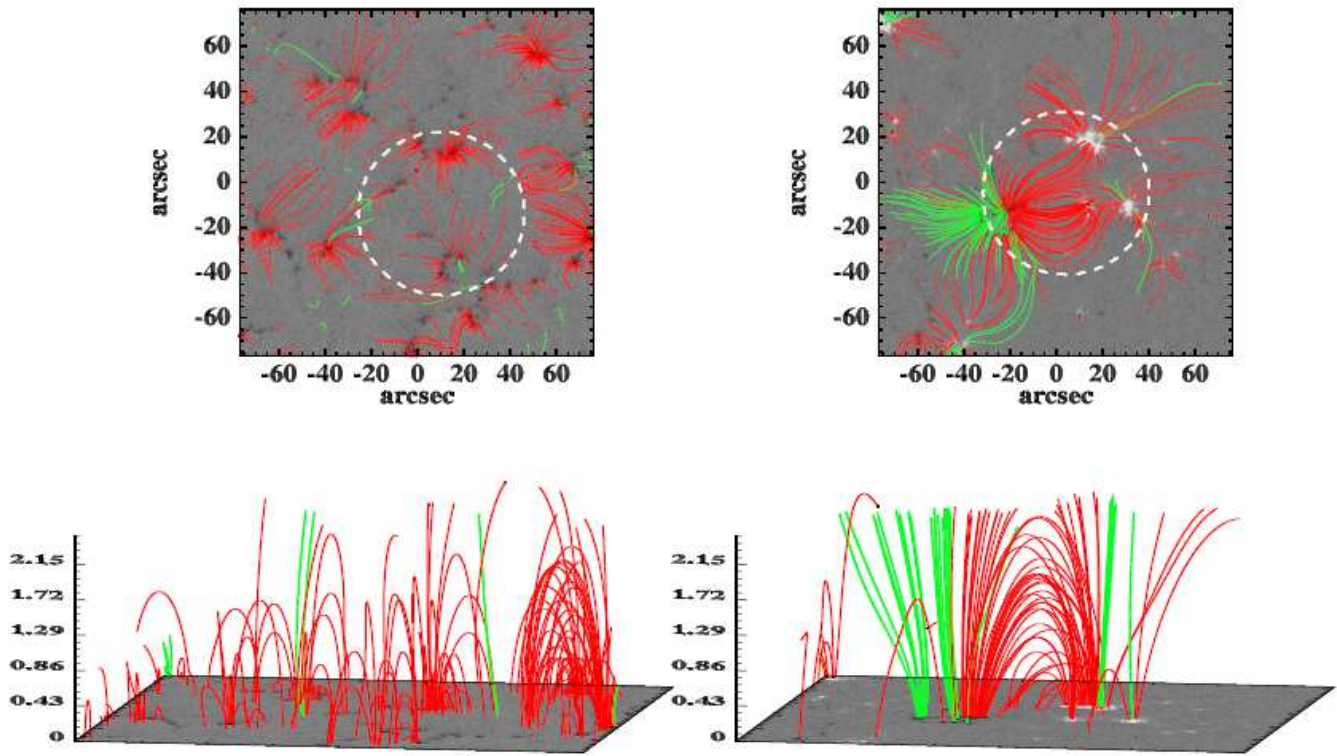}
\caption{Representative field lines from the MDI-extrapolated potential magnetic field. {\it Left 
panels}: data set 1. {\it Right panels}: data set 2. The green thick lines trace the open field lines, 
the red thin line draw  field lines closing within the considered area. The white dashed circles 
delineate the IBIS FOV for each data set. The vertical scale in the bottom panels is the height given in 
Mm.}
\label{field_line}
\end{figure*}

\subsection{The IBIS data sets}\label{s_obs_ibis}
The data utilized in this work were acquired in two quiet areas in close proximity to disk center,  
on 2004 May 31 (in the following data set 1) and 2004 June 02  (data set 2). In both days the 
same acquisition scheme was adopted, sampling sequentially the photospheric \FeI  line at 
709.0 nm and
the  chromospheric \CaII 854.2 nm line.
The time necessary to scan the lines was, respectively, 4 s and 7 s, while the overall 
cadence for the full sequence was 19 s (the Fe\,II 722.4 nm line was also included in the sequence, but not used 
in this paper). The spatial scale was set to 0.166"/pixel ($\approx$ 120 km at the solar surface). 
The photospheric data for set 2 has been utilized in \citet{janssen_06}, while both photospheric 
and chromospheric data, again for set 2, have been analyzed in \citet{noi_07}, hereafter V07. 
The \CaII data for both days has been further described in Paper I. 

We analyzed two stretches of 55 minutes of continuous observations (175 time steps) 
obtained in good to excellent seeing conditions on each day. Data set 2 had a better 
seeing than data set 1. 
We examined the full spectral profiles, as well as studying parameters extracted from them. In particular, 
for each spatial position in the FOV we extract by means of spline interpolation both the 
intensity and the position of the minima of the spectral line profiles. The latter are interpreted in 
terms of Doppler-shift, with the  zero position for the velocity scale defined as the
spatio-temporal average of the whole dataset over a quiet portion of the FOV \citep[see also 
discussion in][and Paper I]{janssen_06}. 

Fig. \ref{fig_fov} gives a synopsis of the data around the middle of the observational sequence 
for both days.  The observed FOV is shown at different wavelengths { together with the simultaneous high resolution (HR) MDI magnetograms.}
The leftmost column shows the broadband continuum at 710 nm, indicating on both days a 
mostly quiet scene, with 
slightly lower contrast in the magnetic regions that delineate 
the supergranular 
network. At higher magnification these can be resolved as tiny bright points embedded within 
the intergranular lanes  (especially in the speckle-reconstructed image of data 
set 2). The second column displays the high resolution  MDI data obtained simultaneously 
to the IBIS data sets: the network elements are well discernible in the two FOVs, but highlight a 
very different magnetic environment, with a weaker, unipolar network in data set 1, and an 
enhanced bipolar network for data set 2. The third column shows  IBIS images acquired at 
about 0.1 nm from the core of the \CaII 854.2 nm line. Together with the reversed granulation 
pattern, these monochromatic images display bright elements with a one-to-one 
correspondence to magnetic elements \citep[][Paper I]{leenaarts_06}. Finally, the fourth column 
shows the images in the core of the \CaII line, that outline a ``segregated'' picture, with part of 
the FOV occupied by fibrils originating from even the smallest magnetic elements, and part of it, 
farthest from the magnetic network, showing an abundance of small bright points surrounded by much 
darker regions (compare the temporal evolution of this region from Movie 2 in Paper I).

\subsection{The magnetic environment}\label{s_obs_mag}

High resolution MDI magnetograms are available for almost the whole duration of the IBIS 
observations on both days. They have a spatial scale of 0.6"/pixel, and a cadence of 1 minute. 
The MDI maps  outline a very different magnetic configuration of the two regions (Fig. \ref
{fig_fov} panels {\it b}, {\it f}). Coronal images confirm  this difference:  while at photospheric 
level both regions could be classified as quiet, EIT Fe XII 19.5 nm images acquired around the 
time of the IBIS observations clarify that the May 31 region is localized at the edge of an 
equatorial coronal hole, while the June 2 region corresponds to a decaying coronal bright point  
(Fig. \ref{fig_foveit}). In neither case did we observe significant variations of the magnetic 
configuration over the duration of the IBIS observations.

The differences in  magnetic topology are already reflected at the chromospheric level, in the 
morphology shown by the \CaII line core intensities. In set 2 we observe a large number of 
elongated fibrilar structures, with a complex spatio-temporal evolution and connecting, in many 
cases, regions of opposite polarity. On the contrary, the number and length of fibrils in data set 
1 are much lower, suggesting either short loops that close nearby the network elements, or more 
vertical structures that follow magnetic field lines extending into interplanetary space.
To assess qualitatively  the 3-D magnetic field configuration, we compute a potential field 
extrapolation starting from the MDI longitudinal field, assumed as the vertical component of the 
vector field at a height of about 200 km. The area utilized for the extrapolation is about six times 
larger than the IBIS FOV. In  Figure \ref{field_line} we show the  resulting magnetic fields, 
extrapolated up to a height of  2.5 Mm. Representative closed lines within this extended region 
are drawn in red (thin lines), while field lines that open to more distant regions or the 
interplanetary space are drawn in green (thick lines). 
The closed field lines closely resemble the chromospheric morphology: in data
set 2 the volume is dominated by field lines connecting  opposite polarity network points, 
spanning large fraction of the FOV, and reaching heights up to about 2 Mm. The quieter data 
set 1, instead, is characterized by short, and low-lying,  closed field lines fanning out from the 
network points and reaching  to nearby internetwork weak magnetic concentrations. The fibrillar 
structures so prominent in the \CaII core images appear a reliable proxy for the presence of a 
magnetic canopy (Paper I, V07).

\begin{figure*}[t]
\centerline{
\includegraphics[scale=1.0]{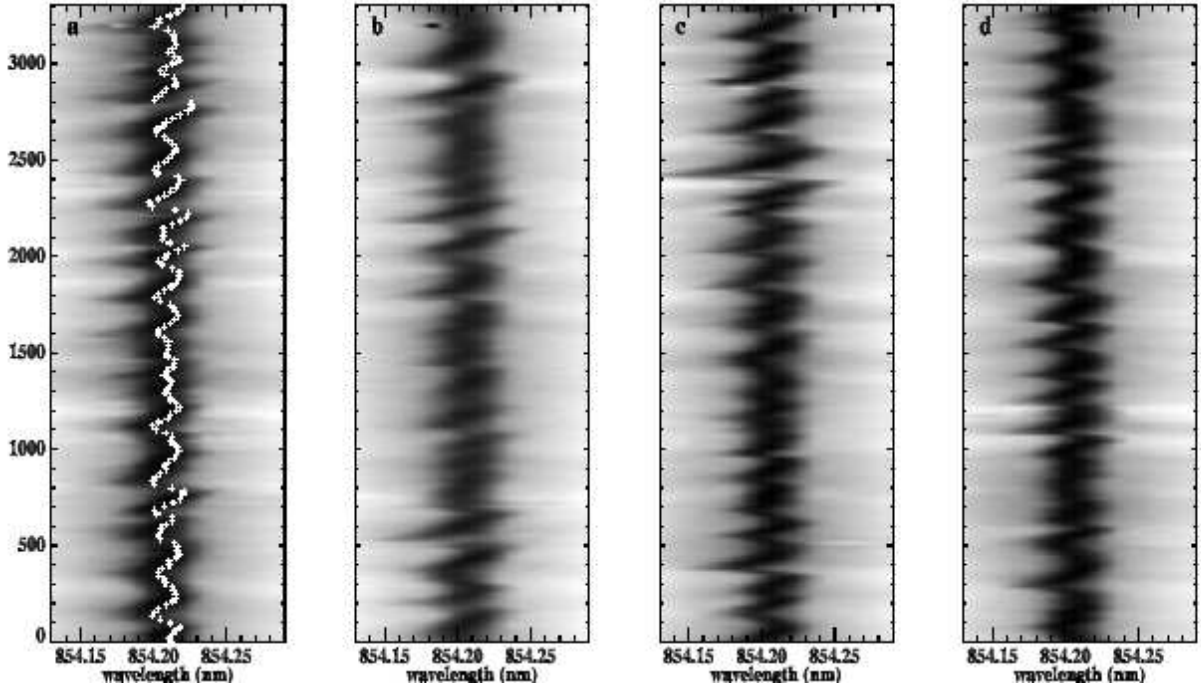}
}
\caption{Ca II 854.2 nm spectral profiles vs. time (given in s), for four positions within the 
internetwork. Panels {\it a, b} refer to data set 1; panels {\it c, d} to data set 2. The time axis 
spans the whole duration of the observations. Note the distinctive saw-tooth appearance. The 
thin white line in panel {\it a} indicates the evolution with time of the line core Doppler shift. 
Maximum velocities reach 6--7 km s$^{-1}$ from average position.}
\label{specline}
\end{figure*}

\begin{figure}

\hbox{
\hglue -1.cm
\includegraphics[scale=0.7]{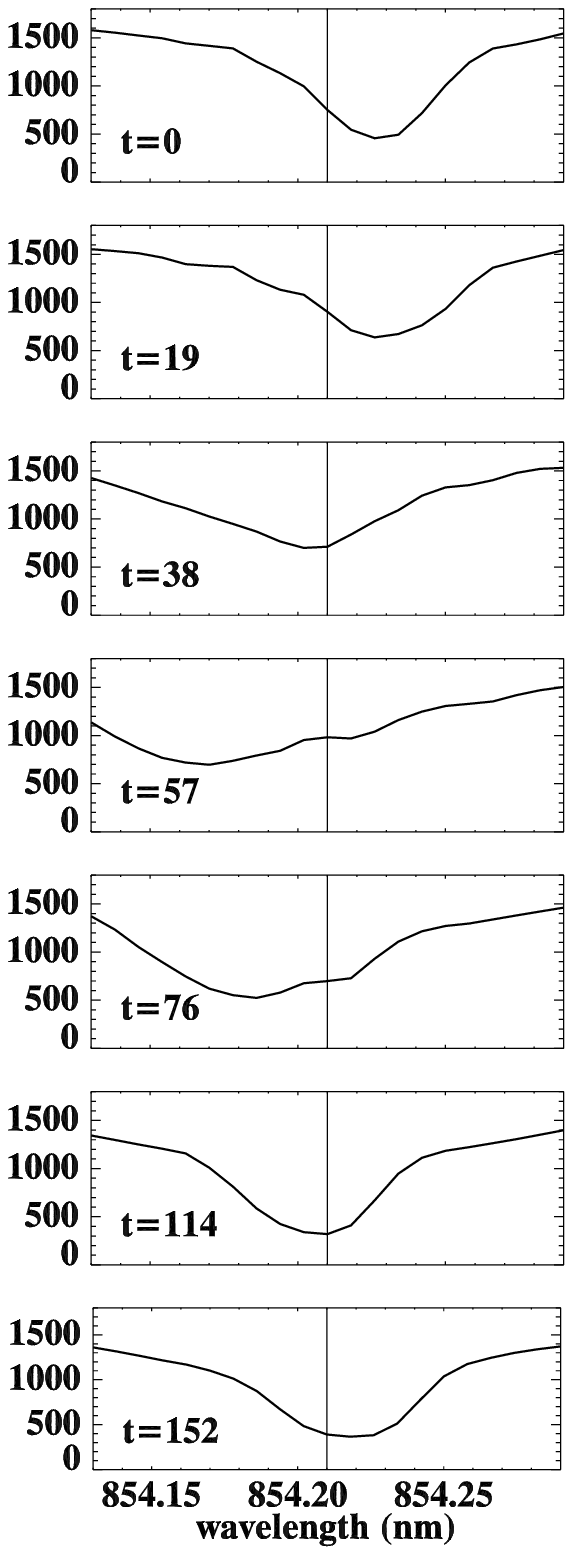}
\hglue -1.55cm
\includegraphics[scale=0.7]{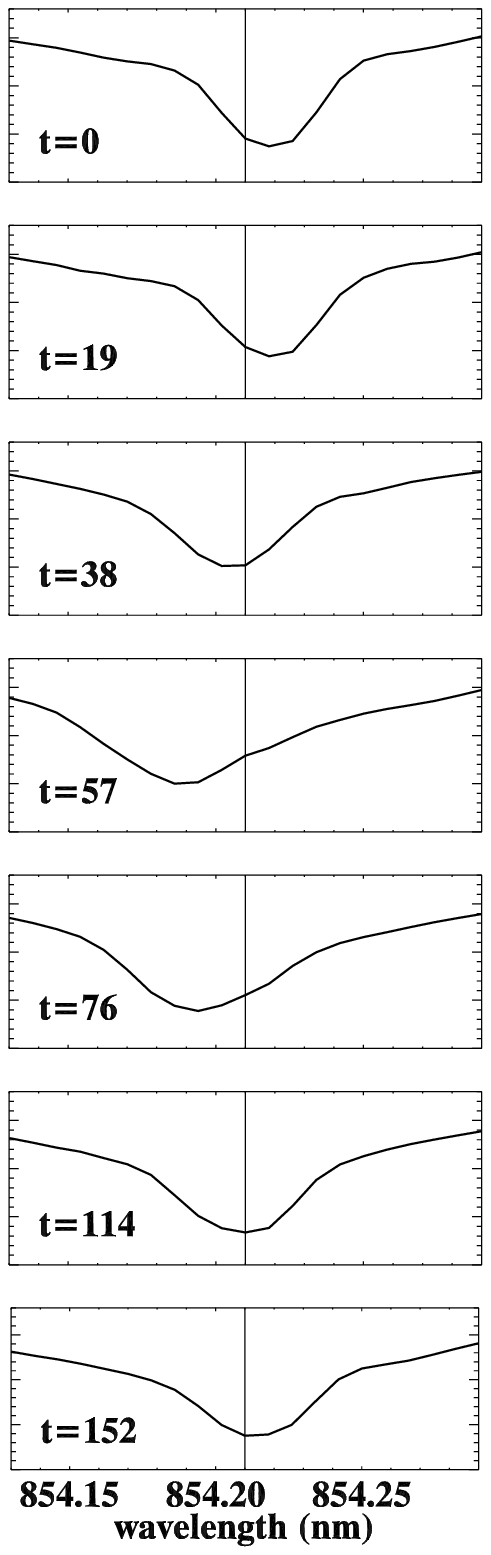}
}
\caption{Temporal evolution of the \CaII 854.2 line during a shock, for two internetwork pixels. 
Intensity is given in arbitrary units. The time indicated in each panel is expressed in s from an 
arbitrary zero point. {The thin vertical line indicates the average position of the line core throughout the full spatio-temporal sample.} Note the broadening of the line during the upward phase.} 
\label{fig_plotshock}
\end{figure}

\section{Evidence for acoustic shocks in the internetwork} \label{s_evid_shocks}

As for the \CaII H and K lines, the formation of \CaII 854.2 nm spans a wide range of 
atmospheric heights (Paper I). From the far wings in toward the core, the line gradually {samples} from the low to
the high photosphere, 
while the line core itself is formed in
the lower chromosphere. However, as shown
by \cite{skartlien_94} and more recently {by} \cite{pietarila_06}, the
presence of hydrodynamic shocks in the non-magnetic chromosphere 
{determines a large range of heights for the formation of the line core, spanning}
from as low as
700 km up to 1300 km. The shocks induce rapid 
variations in the thermodynamics of the affected atmosphere, {resulting} in asymmetric 
\CaII 854.2 spectral profiles \citep[cf. Figs. 3 and 7 of ][]{pietarila_06}.

Fig. \ref{specline} displays the observed temporal evolution of the IBIS \CaII 854.2 nm spectral 
profiles for four representative pixels, two for each data set. The pixels have been chosen as 
typical of the internetwork {regions} of the FOV.
As described in Paper I, the sequence of line core images shows a rapidly evolving scene with 
areas that are seething with small bright features immersed in an otherwise dark background 
(compare Movie 2 in Paper I). The spectral profiles of Fig.  \ref{specline} clearly show how this 
appearance is the result of the line being repeatedly shifted, strongly towards the 
blue followed by a slower, smaller amplitude drift back toward the red. The pattern is repeated at the 
dominant chromospheric periodicity of $\approx$120--180 s throughout the whole course of the 
observations. Such strong line shifts are viewed as 
an enhanced brightness when observed in the narrowband, fixed-wavelength core images; 
their spatial coherence determines how the bright features are observed with respect to the 
background intensity  (see Section \ref{s_properties}).

A non-linear steepening of the acoustic waves propagating upward from the photosphere
at periodicities shorter than the acoustic cut-off period of $\approx$180 s is prescribed by the 
strong decrease in density and the approximate conservation of wave energy. This 
is clearly reflected in the spectra of Fig. \ref{specline}, where we observe both a deviation of the 
oscillations from a sinusoidal form, and a large amplitude of the excursions. At these 
periodicities, the typical r.m.s. of chromospheric Doppler shifts is around 1 km s$^{-1}$, vs. $
\sim$0.07 km s$^{-1}$ of photospheric motions,  cf. \citet{reardon_turbulence_08}. 
However, many of the line shift episodes not only display much stronger Doppler amplitudes, of 
the order of 5-6 km s$^{-1}$, 
but also exhibit a very abrupt change from large redshift to large blueshift, often within a single 
19-second temporal step in our sampling. Further,
the maximum redshift of the line often coincides with a brightening episode just blueward of the line core, 
lasting 50--70 s,  with enhanced 
intensities up to 1.5--2  times the average value at this wavelength.  They are also 
associated with an increased intensity in the extended line wings, that progresses in time from 
the far wings towards the core. For the wavelength range of our observations, this 
time is less than about 60 s. Clear examples in Fig. \ref{specline} occur at t=1850 s in panel {\it 
a}, or at t=1050, 1200 s in panel {\it d}. 

These {features} are analogous with those of the \CaII K$_{2V}$ grains \citep[][RU91]
{liu_74,cram_77,cram_83} which have been unequivocally 
identified as due to hydrodynamic shocks in the mid chromosphere (hereafter we will use the 
terms grains and shocks interchangeably). A clear example of the resemblance between the 
temporal evolution of the \CaII K and \CaII 854.2 lines in a non-magnetic atmosphere is 
provided by comparing Fig. \ref{specline} with Fig. 2 of \citet{kamio_03}: all the 
characteristics previously described for the IBIS data appear very obvious in the latter 
high-resolution observations (note the different axes of their  Figure with respect to Fig. \ref
{specline}). The Doppler shifts are clearer in the \CaII 854.2 line while the bright grains and 
related wing enhancements are less prominent when compared to the H and K 
lines. These differences are a 
consequence of the different formation and wavelengths of the two lines: the 
\CaII 854.2 line has a much stronger Doppler sensitivity due to its longer wavelength 
and the Planck function (to which the emissivity is related) has a much weaker 
temperature dependence in the near-IR. 

To our knowledge, this is the first time that acoustic shocks in the quiet internetwork 
chromosphere are convincingly (and abundantly) observed in the \CaII 854.2 line, due 
to both the high spectral and spatial resolution of the IBIS data.
Indeed, synthetic profiles  calculated with the 1-D hydrodynamical code of CS97 well agree with 
the features described above \citep[Pietarila, private communication;][]{cauzzi_07}. As further 
evidence, we show in Fig. \ref{fig_plotshock}  the development of \CaII 854.2 spectra of two {example}
pixels, at several times during the passing of a shock wave. {The profiles in the right panels are representative of a large fraction of  internetwork pixels, for which the characteristic shifts and asymmetries}
of the spectral line are in excellent agreement with the results of the simulations displayed 
in Fig.7 of \citet{pietarila_06}. {As an interesting case, we also show  in the left panels the spectral profiles for a more extreme event, where a strong shock greatly deforms the whole line shape up to the far blue wing.} 
 
\begin{figure*}
\centerline{
\includegraphics[scale=1.0]{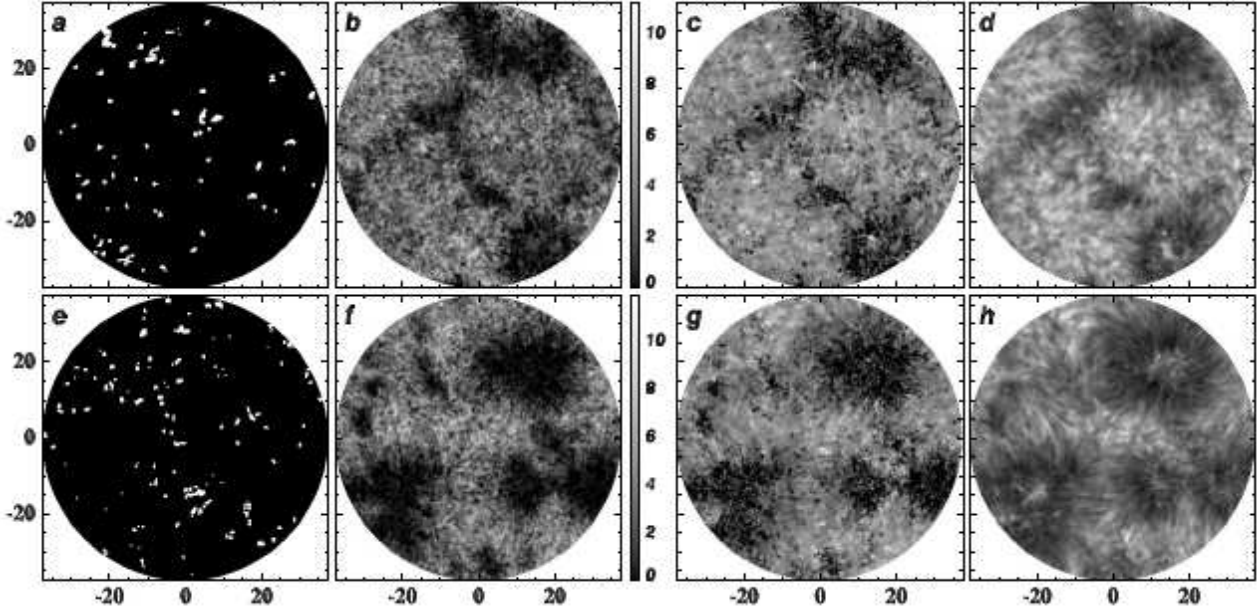}
}
\caption{Spatial distribution of properties related to shocks. Top row: data set 1; bottom row: 
data set 2. Results shown in {\it a} through {\it c}, and {\it e} through {\it g} have been derived 
from the POD analysis. Axes are in arcsec.  Panels {\it a, e}: binary ``shock maps'' at a time 
around the  middle of the sequence. Panels {\it b, f}: cumulative number of shocks within the 
whole observational sequence. Panels {\it c, g}:  maps of the shocks' cumulative amplitude over 
the whole duration of the observations. Panels {\it d, h}: spatial distribution of the Fourier power 
for chromospheric velocity, integrated over frequencies at and slightly above the acoustic cut-
off (5.5 -- 8.0 mHz).}
\label{map1}
\end{figure*}

\section{Shock identification}\label{s_findshocks}

Typically, shocks in extended FOV have been identified by using an intensity threshold 
on  \CaII H$_{2V}$ or K$_{2V}$ filtergrams. This takes advantage of the larger
temperature sensitivity of the Planck function in the blue. However,
shocks are foremost a velocity phenomena, due to the steepening and rapid 
reversal of the upward propagating waves. Hence, the velocity signature 
is presumably the most suitable means to 
identify their actual occurrence. Given the 
properties reminded above, 
\CaII 854.2 appears to be a
reliable line with which to identify shocks in the solar chromosphere. 

We adopt two 
different methods that utilize the temporal sequence of the full spectral 
profiles to locate shocks in our datasets.
For clarity of exposition, we shortly describe here only their most relevant characteristics,
with further details given in the Appendix.
The first method we call {it} the ``velocity'' method: analyzing the temporal profile of the 
chromospheric line-of-sight velocity for each spatial position, we identify times of brusque, 
strong displacements from downward to upward motions. The method utilizes a threshold both 
in amplitude of the displacement, and in the temporal interval in which this displacement occurs, 
set at a maximum of two time steps of the IBIS sequence  (i.e. 38 s).
The second method uses the ``Proper Orthogonal Decomposition'' technique \citep[POD,][]
{lum_96}, that separates a time-series of spectral profiles into multiple orthogonal eigenmodes. 
These are then ordered in terms of decreasing energy content, i.e. of their 
contribution to the whole intensity profile. The periodic velocity shifts that produce the 
sawtooth shape in the temporal sequence of internetwork \CaII 854.2 spectra is clearly 
identified by the POD in a single antisymmetric 
eigenmode (see panel {\it b} of Fig. \ref{eigf_pod}), properly modulated by a temporal 
coefficient. After finding all 
the spatial positions for which this eigenmode represents the dominant mode,
we define the shock occurrence at the times when the positive peak of the fluctuation shifts 
from the red to the blue side of the line. We apply the same temporal threshold as in the 
velocity method, but no amplitude threshold.

For both methods we derive a time series of binary ``shock maps'' defining the times of occurrence and spatial distribution of the 
shocks. An example of the POD results is given in Fig. \ref{map1} {\it a} and {\it e}. 
The results from the two methods are similar. As a general {characteristic}, the POD 
identifies smaller shocks occurring more frequently, most notably in the areas filled with fibrils, 
for which the velocity method essentially finds no shocks (see Sects. \ref{ss_size} and \ref
{ss_spatial} below). This can be attributed to the absence of a prescribed threshold for the 
amplitude of the shocks in the POD case. We will point out other differences and  similarities 
as we describe the further results of the analysis.

\section{Shock properties} \label{s_properties}

\subsection{Spatial size and occurrence}\label{ss_size}

From the shocks maps defined above,  we can derive the size of a shock by  
measuring the area covered by contiguous pixels possessing values of one. 
We discarded any shock {area} covering less than 2 pixels in either direction, imposing an 
effective lower limit of 0.3 $\times$ 0.3 arcsec$^2$ to their dimension. For both data sets we find 
average shock extents of 0.4 arcsec$^2$ (using the POD method) to 0.8 arcsec$^2$. Under the
assumption that the shocks are roughly circular, this corresponds to diameters of 0.7 arcsec to 1.0 arcsec.
These values can be compared to the typical 
1-2 arcsec size of the 
K$_{2V}$ grains as reported in the literature \citep[e.g. RU91;][]{liu_74,tritschler_07}, although 
a number of factors set apart our analysis from the more common intensity threshold approach.  

The average number of  shocks  occurring at any time within a given area can also be 
estimated. 
To this end, we selected for both data sets a 30'' diameter area roughly covering 
the center of a supergranulation cell centered at $(x,y) = (14,0)$ 
for data set 1 and ($(x,y) = (-4,2)$) for data set 2 (see Fig. \ref{fig_fov}).
We find an average number of shocks within these areas ranging from 13 (velocity 
method, data set 2) to 24 (either method, data set 1). 
The number of K$_{2V}$ bright grains in a typical supergranulation cell have been reported at 
between
10 and 20  in ``the best K$_{2V}$ spectroheliograms'' (RU91). Our average values are only 
somewhat higher than these previous findings, but we observe large variations of the number 
of shocks during the course of the observations, mostly correlated to the seeing conditions: 
during moments of good to excellent seeing, more than 30 shocks can be identified within a 
single cell.

\subsection{Temporal recurrence}\label{ss_temporal}

We can further analyze the temporal behavior of the shock occurrence {by using two methods. The first one, described
in the Appendix, 
shows that the spatial areas where shocks occur in large numbers are dominated by 
shock periodicities around the three minutes typical of the
chromosphere}.
We also take another approach here by calculating the histograms of the ``waiting times'' 
for the shocks, defined simply as the interval between one shock and the next. No significant 
difference is found between the two methods of shock identification.

\begin{figure}
\centerline{
\includegraphics[scale=1.0]{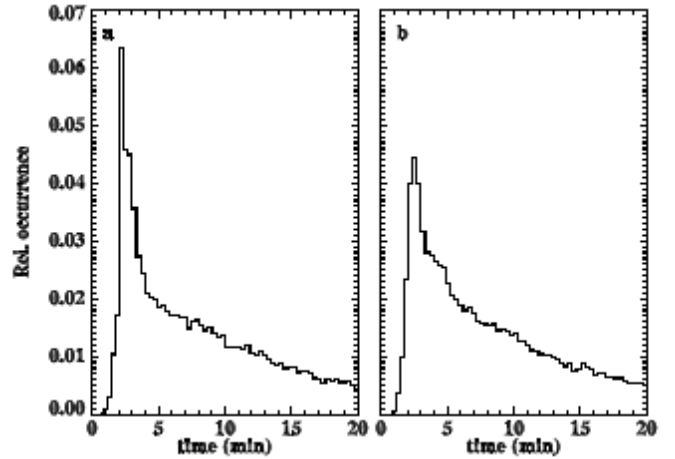}
}
\caption{Distribution of waiting times between shocks, for data set 1 (panel {\it a}) and data 
set 2 (panel {\it b}).}
\label{fig_wait_time}
\end{figure}

The results {are} shown in Fig. \ref{fig_wait_time} where a strong peak around 120--150 seconds 
is observed for both data sets, with a sharp decline at 
shorter time scales and a slower decrease towards longer periods. The data do not show 
any evidence of shock periodicities below 120 s (i.e. frequencies above 8 mHz), although such 
periods would be observable with our data. The long tail of the curves at longer periodicities highlights 
the fact that \CaII 854.2 shocks do not occur in a continuous fashion, but often in separate 
bursts. At the same time, no obvious values of waiting time between bursts emerge from the 
curves, although a change in the slope of the distribution appears around 4--5 minutes.
Assuming this values as discriminant between shocks in the same burst, and between different 
bursts, we find that the average number of shocks per burst is only 1.25, and that the average 
number of bursts within the whole sequence is around 3. 

All of this bears on the total number of shocks observed in each pixel,  displayed in panels  {\it 
b, f} of Fig. \ref{map1} (using the POD results). From these maps  it is clear that most of the 
internetwork locations only develop shock for a fraction of the observing sequence: only between 25\%
and 30\% of the ``shocking pixels'' display 5 or more shocks during the whole sequence. The 
maximum number of shocks for any given pixel is around 15 for both data sets and methods, 
although this is attained only in a very few pixels.

\subsection{Spatial pattern} \label{ss_spatial}

A striking result emerging from panels {\it b, f} in Fig. \ref{map1} is the extreme inhomogeneity  
of the spatial
distribution of the total number of shocks.  In particular, large portions of the internetwork 
clustered around the magnetic elements 
experience very few shocks, or none at all. The velocity method provides the same spatial 
patterns, but with even more extended regions displaying {very few shocks} . In the maps of Fig. \ref{map1}, 
more than 20\%  of the pixels in data set 1, and 30\% in data set 2, do not experience any 
shock  throughout the whole course of the observations. These values increase to a 
surprising 50\%, respectively 60\%, for areas with a total number of shocks $\le 2$. 
These areas are significantly larger {than} the area occupied by the photospheric magnetic elements, 
less than 10\% in both cases. These results highlight how chromospheric dynamical 
properties are poorly constrained by the usually adopted photospheric magnetic diagnostics, a 
fact already remarked in Paper I.

We further note that these {``very few shocks''} areas are essentially coincident with the fibrils 
observed in the \CaII line core images (compare Fig. \ref{map1} with Fig. \ref{fig_fov}), that 
are again revealed as a crucial player in shaping the chromospheric dynamics (cf. V07 and Paper 
I).
Conversely, the highest number of shocks is recorded in areas well removed from both the 
photospheric magnetic network {\it and} the fibrils.  These areas are consolidated in small patches 
scattered within the internetwork.

\begin{figure}
\centerline{
\includegraphics[scale=1.0]{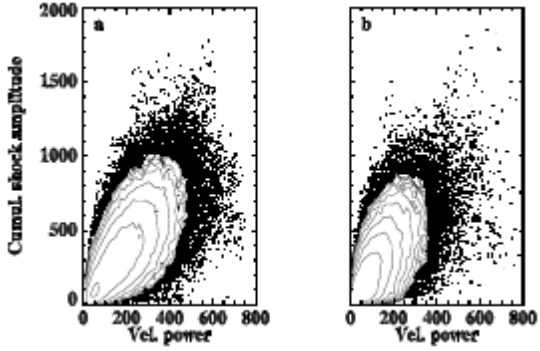}
}
\caption{Scatterplot of the cumulative amplitude of shocks vs. Fourier power of chromospheric 
velocities, integrated in a band of frequencies just above the acoustic cut-off (5.5-- 8.0 mHz). 
Both axes are in arbitrary units. To avoid pixel crowding, the 2-D distribution is represented by contours, in logarithmic levels.}
\label{f_scatter}
\end{figure}

\subsection{Shocks and Fourier velocity power} \label{ss_power}

Panels {\it c, g} of Fig. \ref{map1} provide the cumulative amplitude of the shocks over the 
duration of the observations, again derived from the POD analysis (Sect. \ref{ss_pod}).

It is interesting to compare the  amplitude maps with the distribution of the chromospheric 
velocity Fourier power,  integrated over the 5.5-8.0 mHz range just 
above the acoustic cut-off frequency.
The latter is shown in panels {\it d, h} of Fig. \ref{map1}.
As discussed in V07 for a subset of the data 
utilized here,  in such maps all the magnetic elements are surrounded by areas of very low 
velocity power, corresponding to the ``magnetic shadows'' first introduced by \citet{judge_01} in 
an analysis of chromospheric SUMER data.  They are also coincident with the fibrillar structures 
so prominent in \CaII core images and, as clear from Fig. \ref{map1},  with the chromospheric 
regions where few shocks occur during the observational sequence.

The scatterplots in Fig. \ref{f_scatter} display a clear pixel by pixel relationship between the 
velocity power and  the cumulative shocks amplitude: a simple linear regression finds a 
correlation of about 0.65. From these {correlations}, we surmise that a large fraction of what is commonly 
identified  as the ``3-minute chromospheric oscillations'' signal in Fourier analysis \citep[e.g.][]
{orrall_66,noyes_67,deubner_90} is  due to the presence of shocks, even if they do occur with the 
intermittent character described above. {In practice, } the linear Fourier analysis identifies the shock 
periodicity and provides a measure of the amplitude of the velocity fluctuations at this dominant 
frequency, even though the physical phenomenon that produces this signal is very non linear.

\begin{SCfigure*}
\centering
\includegraphics[scale=1.0]{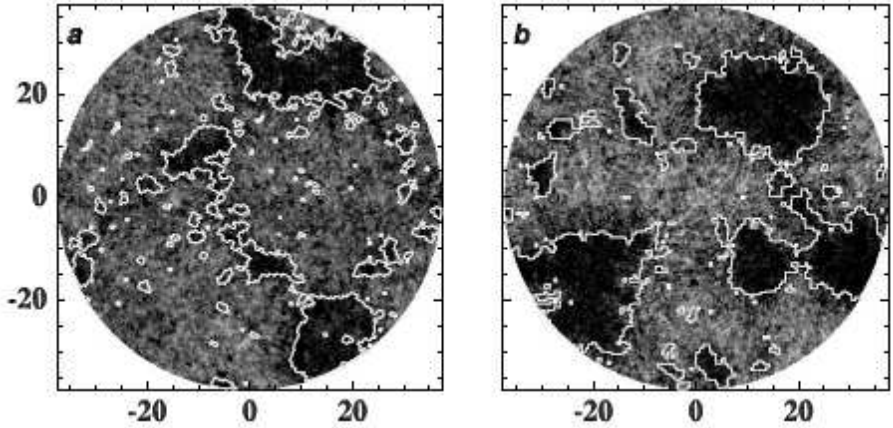}
\caption{Cumulative shock maps for data set 1 (panel {\it a}) and 2 (panel {\it b}). Overlaid is the contour of the  
high resolution MDI magnetic flux, averaged over the course of the IBIS observations. The 
contour level is set at 8 G. Compare how even minute magnetic structures within the 
internetwork, probably occurring only for a fraction of the observations, correspond to a 
decreased number of  \CaII shocks. Spatial scale in arcsec.}
\label{f_mdi_shocks}
\end{SCfigure*}


\section{Shock occurrence vs. magnetic structures}\label{s_mag}

Numerous observational studies have searched for a connection between the occurrence of K$_
{2V}$ grains and the presence of  magnetic structures in the internetwork as evidence for a possible 
excitation mechanism \citep[see e.g. the Introduction of][]{lites_99}. The majority of these have 
concluded that the grains are essentially a hydrodynamical phenomenon, for which 
magnetism plays a minor role at most. However, there remain some claims about a tight 
correspondence between the presence of small scale bipoles in the internetwork and  K$_{2V}$ 
grains, most recently by \citet{sivaraman_00}.  
Our comprehensive data sets,  combining high-cadence spectral data over extended areas with 
actual magnetic field measurements, can be used to examine this question.

\begin{figure}
\centerline{
\includegraphics[scale=1.0]{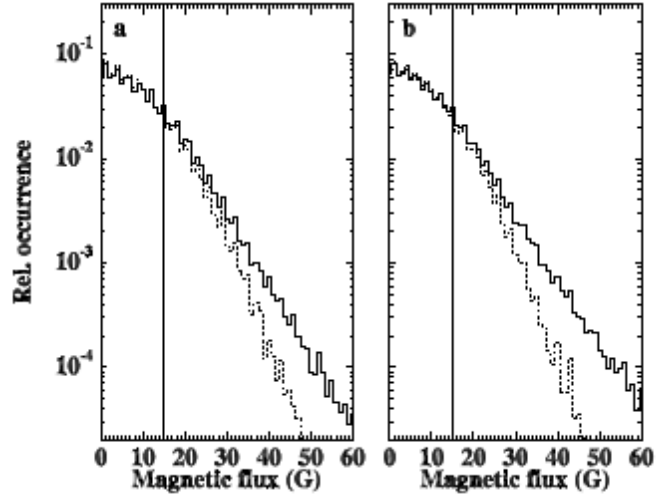}
}
\caption{Distribution of MDI longitudinal magnetic flux measured in points undergoing a large 
(dotted line) and small (solid) number of shocks, respectively. }
\label{f_distr_b}
\end{figure}

A coarse inspection of the magnetic maps of Fig. \ref{fig_fov} and the cumulative shock maps 
of Fig. \ref{map1} immediately shows that shocks do {\it not} occur at the locations of magnetic 
structures. A closer inspection shows that the shocks apparently avoid 
even the smallest magnetic elements: in Fig. \ref{f_mdi_shocks} we show again the 
cumulative shocks maps for both data sets, overlaid with the contours of the time-averaged 
MDI magnetic maps. {While in the average MDI maps the noise is reduced to a level of about 5 G,} in Fig. \ref{f_mdi_shocks} the magnetic contour level is set at 8 G 
to allow visibility of the weak and/or transient structures (the network elements have an average 
magnetic flux of 200--300 G).
>From {Fig. \ref{f_mdi_shocks}, one infers} that even small magnetic structures impede the development of 
the shocks: one notes the small ``shadows'' in 
the shock maps surrounding several tiny internetwork magnetic structures, for example at 
positions ($-$30,10); ($-$10,$-$7); (11,2) in data set 1, or ($-$14,27); (12,$-$33) in data set 2.

{A deeper analysis on} this issue involves measuring the magnitude of the magnetic field in 
those areas that undergo shocks. 
{We first selected, in both data sets, the quiet regions where} the temporal 
average of the MDI magnetic flux does not exceed 30 G.  We then divided these regions in 
two parts, one collecting the pixels {where several shocks ($n \ge 5$), over the duration of the 
observations, are observed} and the other collecting pixels with a smaller number of shocks ($1\le n\le 
4$). Excluding the $n = 0$ pixels effectively avoids most of the magnetic shadows 
associated with the network elements (cf. Fig. \ref{map1}). The magnetic flux distributions 
obtained from the temporally resolved MDI maps in the two classes of pixels are shown in Fig.  
\ref{f_distr_b}. The vertical line at 15 G represents the noise level for a single pixel in MDI 
maps. In both data sets, the distributions for the two classes of pixels start diverging shortly 
after the noise value, with the pixels undergoing many shocks characterized by sensibly lower 
values of the magnetic flux.

The curves diverge at the level of $10^{-2}$ in relative occurrence, i.e. only a small fraction of 
the pixels displays this effect. Nevertheless, the difference is clear. Combined with the 
maps in Fig. \ref{f_mdi_shocks}, this shows that the internetwork magnetic 
fields actually {seem to have} a {\it negative} effect on the development of shocks.
Magnetograms with higher sensitivity and spatial resolution than those of MDI might 
reveal if there is a critical area or flux value below which this effect no longer holds.

\section{Shock occurrence vs. photospheric velocities} \label{s_phot}

We now examine the {relationship} between the occurrence of shocks and the underlying
photospheric dynamics. We do so both with classical Fourier analysis, and also using
wavelet analysis on photospheric velocities, which is a more efficient way to take into account the 
highly intermittent character of the shocks. 

\subsection{Fourier phase difference spectra}\label{ss_phase}

The spatially-averaged Fourier velocity power spectra for these data sets are displayed in \citet
{reardon_turbulence_08} and Paper I, and are in general very similar to previously published 
spectra \citep[e.g.][]{deubner_90}:  the photospheric curve shows a strong maximum around 
the classical $5$ minute periodicity ($\nu \sim 3.5$ mHz), while the chromospheric curve for
the internetwork shows a 
flatter maximum from $\sim 4.5$ to 7 mHz. Here we concentrate instead on the phase relationship 
between the photospheric and chromospheric velocities as a means to study 
the propagation of acoustic waves.

\begin{figure}
\centerline{
\includegraphics[scale=1.0]{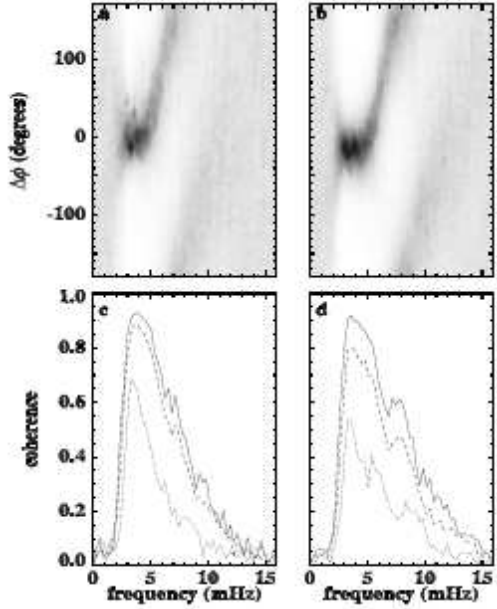}
}
\caption{Panels {\it a, b}: Fourier spectra of  phase differences between the photospheric and 
chromospheric velocities, for data set 1 and 2, respectively. The diagram has been calculated 
over the points in the FOV that develop at least 5 shocks throughout the observational 
sequence. Panels {\it c,d}: corresponding coherence spectra (solid line). Dashed lines refer to 
the coherence measured for an equally populated sample of pixels undergoing less than 5 
shocks; dotted lines to the coherence for the case of fibrils.}
\label{phase}
\end{figure}

Panels {\it a, b} in Fig.  \ref{phase} display the phase difference spectra between the Fourier 
transforms of the chromospheric and photospheric velocities, obtained over the pixels that 
develop at least  5 shocks throughout the observational sequence.  
The Figure is built by plotting the binned  phase difference $\Delta\phi$ weighted by the cross-power amplitude and normalized per temporal frequency bin; for details see \cite{kri_01} and 
references therein. The noisier scatterplot for data set 1 is  due to the worse seeing conditions 
with respect to data set 2. A positive phase difference describes signals 
propagating upward in the solar atmosphere: the scatterplots thus clearly evidence the 
propagation of acoustic waves from photosphere to chromosphere, up to about 10--12 mHz.
The bottom panels  {\it c, d} provide the coherence spectra of the Fourier velocity signal (solid 
line). The coherence is very high around the velocity power peak at  5--6 mHz (see Paper I), 
and falls below the confidence limit of 0.5 around 8--9 mHz.

If we consider an equal-size sample of pixels characterized by a smaller number of shocks, we 
obtain essentially the same phase differences (not shown in Figure) and slightly lower  
coherence values (dashed line):  the same photospheric piston mechanism thus seems at the 
base of all quiet chromospheric oscillations, regardless of whether other atmospheric properties 
are conducive to the development of many mid-chromospheric shocks or not. The coherence is 
instead much smaller if we use an equal sample of pixels belonging to the fibrillar areas (dotted 
line); these are areas
where most pixels do not develop any chromospheric shocks, especially for  data set 2. 
In these atmospheric regions, a direct (vertical) relationship between photospheric and 
chromospheric dynamics no longer holds.

The coherence spectra display a curious bump around 7 mHz, most evident in data set 
2. Such a feature has already been reported by \cite{deubner_90} in a spectrographic analysis 
of a quiet region. Like these authors, we find no obvious explanation for the feature. 
We suspect that it might be related to the "aureoles" of  enhanced high frequency power 
sometimes measured in both photospheric and chromospheric signatures around active 
regions or strong network elements \citep[see e.g.][and references therein]{kri_01}. The bump 
is indeed more evident in data set 2, which has stronger magnetic fields and a more "active 
region-like" magnetic configuration than data set 1.

\subsection{Photospheric velocity power and shock occurrence}\label{ss_wavelets}

In their analysis of K$_{2V}$ grain formation, CS97 remarked that photospheric acoustic waves 
at or near the cut-off frequency play the most important role for the development of shocks. In 
particular, they stated that whenever photospheric velocities have significant power around 5 
mHz, there will usually be significant bright grains. 
To check these statements, we performed a wavelet analysis of the photospheric velocities. 
We concentrate on the spatio-temporal features showing enhanced power at periodicities 
between 120 and 200 s (5--8 mHz), taken as values larger than the average over the whole 
sample  (this accounts for about 35\% of the total number of pixels). Hereinafter we refer to them 
as ``areas'',  even if they contain a temporal dimension. 

\begin{figure*}
\centering
\includegraphics[scale=1.0]{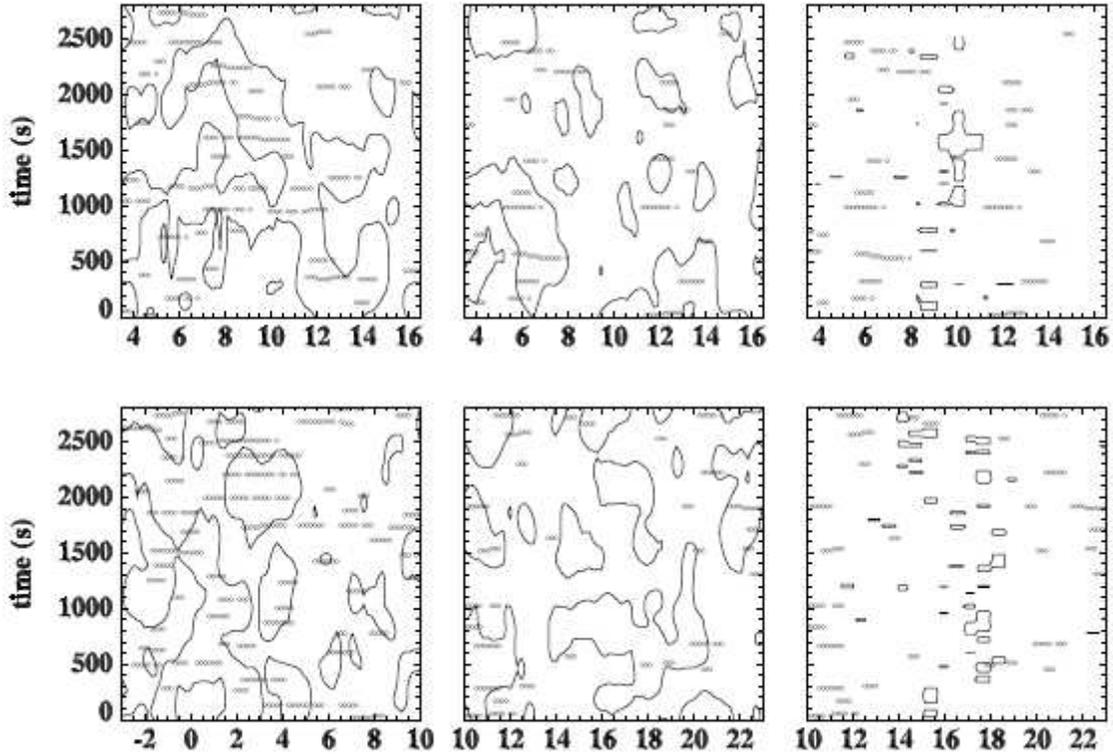}
\caption{
Examples of spatio-temporal occurrence of
chromospheric shocks (small diamonds) in relation to the photospheric 
velocity power at periodicities between 120 and 200 s (left and center panels) {and magnetic features from MDI (right panels)}. Velocity power contour levels
are set at the average value over the whole FOV plus half standard deviation. Top row refers to 
data set 1; bottom row to data set 2. The temporal axis is shorter than the full IBIS observational sequence as wavelet maps are not reliable in the initial and final temporal samples, and because of the delay applied between photospheric  and chromospheric signal.
The  {\it x} -- axes coordinates are given in the same units as in the FOV of Figs. \ref{fig_fov}, \ref{map1}, \ref{f_mdi_shocks}. 
Left panels correspond to {\it y}= 5, 0 for data set 1 and data set 2, and define quiet areas with strong correspondence between photospheric dynamics and shocks' occurrence. Center panels correspond to  
 {\it y}= 12, 3 for data set 1 and data set 2,
and show areas with comparable photospheric dynamics but notably reduced development of shocks. This is due to the presence of small-scale magnetic structures, as outlined by the contours in the right panels.  Contours are set at 30 G for the magnetic flux. For data set 1 the MDI sequence ends before the respective IBIS temporal sequence  (cf. upper right panel).
 }

\label{fig_wavelet_maps}
\end{figure*}

{Fig. \ref{fig_wavelet_maps} shows example that the mechanism outlined by CS97 is most probably at work}. In 
all panels the {\it x}-axis represents the spatial dimension along a horizontal cut within a quiet 
region in the FOV, while the {\it y}-axis displays the temporal dimension over the duration of the 
observations. Top panels refer to instances in data set 1; bottom panels to data set 2. The small 
diamonds represent the occurrence of \CaII 854.2  shocks, while the contours outline the areas 
of enhanced photospheric velocity power  (left and center panels), and the presence of magnetic 
features (right panels). In the left panels, 
most of the patches of strong photospheric velocity power
encompass the areas where shocks are observed, indicating a relationship 
between these two phenomena. The coincidence in the case of data set 1 is 
particularly striking, with photospheric contours enclosing the chromospheric shocks  for the 
whole sequence over extended areas. We note that in order to obtain this coincidence a 
delay of about 120 seconds (6 time steps) has been applied between the photospheric signal and the 
occurrence of shocks. This is consistent with the idea that the photospheric perturbations 
propagate upward at the speed of sound over a $\sim$ 700--1000 km height difference. 

However, determining whether all areas of enhanced photospheric power indeed give rise to 
shocks, proved a difficult task. As the presence of magnetic fields breaks the direct  relationship 
between the photospheric and chromospheric dynamics, one has to carefully select the areas 
to scrutinize this phenomenon. The center panels of Fig. \ref{fig_wavelet_maps} provide a 
clear example of this. The plots have been obtained for 
spatial positions  still well removed from 
the magnetic network, and in them we observe that in several instances, especially for the data set 2,
 areas of enhanced 
photospheric power do not produce any shocks, contrary to the expectations of the 
hydrodynamical modeling. The small magnetic structures responsible for this lack of shocks are outlined in the
right panels, at a level of $\sim$ 30 G: it is obvious how the shocks {generally} avoid both the magnetic elements proper, 
and some surface around them, as for the larger network shadows.

We hence attempted to identify the quietest portions of the FOV for both data sets, by 
combining the series of MDI maps with the \CaII 854.2 core intensity maps (see also Fig. \ref
{fig_three_chr}). In both cases we selected a 8''$\times$13''  area (50$\times$80 spatial pixels, 
considered independent) positioned almost at the center of the FOV, that displayed the least 
magnetic flux in MDI maps and no obvious evidence of fibrils in \CaII core images.
For  these regions, we find that  more than 80\% of the areas with 
enhanced photospheric velocity power contain shocks. Further, the number of shocks 
measured in each area is directly proportional to the spatio-temporal extent of that area: the 
development of acoustic shocks  appears to be continuos as long as photospheric 
conditions are favorable. 
As a rough estimate of the influence of magnetic structures, we compare the total number of 
shocks measured in these quietest portions of the FOV, taken as prototypes for the undisturbed 
acoustic shocks mechanism, with those measured in the whole supergranular cells defined in 
Sect. \ref{ss_size}.  We find that the 
quietest regions produce, per unit area, from almost 2 (data set 1) up to  3 (data set 2) times 
more shocks than the supergranular cells as a whole. 

\begin{figure}
\centerline{
\includegraphics[scale=1.0]{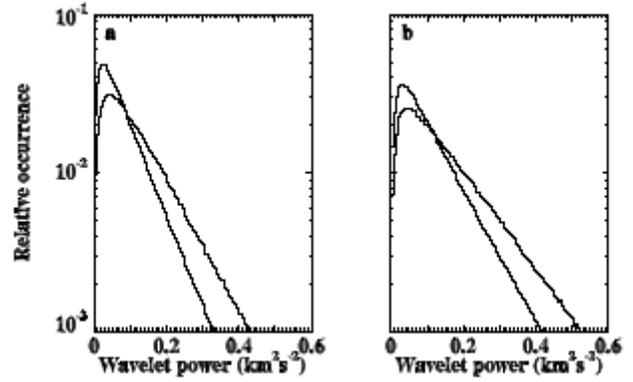}
}
\caption{Distribution of photospheric velocity power between 120 and 200 s periodicity for 
positions undergoing shocks (thick curves) and not (thin curves). Panel {\it a} refers to data set 1; panel {\it b} to data set 2.
The overall higher power values in 
data set 2 are due to better seeing conditions. }
\label{fig_hist_pow5}
\end{figure}

We have shown that high photospheric velocity power at the cut-off periodicity is  certainly 
responsible for the development of shocks. But do all shocks occur in areas of high power?  To 
answer this question, we show in Fig. \ref{fig_hist_pow5}   the 
distribution of  the photospheric velocity power for both data sets. 
The curves have been calculated for all the pixels 
undergoing shocks (thick line) and pixels where shocks are not observed (thin line). Again a 
delay of 120 seconds has been applied between photospheric and chromospheric signals. The 
photospheric power is obviously stronger in the former case than in the latter,  as expected 
from the results described above, with an average value 45\% larger for data set 1, and 35\% for 
data set 2. By measuring the area enclosed below the curves, we find that in both data sets over 55\% 
of shocks occur in areas with photospheric power larger than the average value (the latter 
accounting for 35\% of the total pixels, as stated  before). 
This is to be considered a lower
limit, as  shocks that are part of a common pattern sometimes lie just outside of the areas defined 
by the photospheric level, as clearly seen in Fig. \ref{fig_wavelet_maps}.
Thus, a large fraction of shocks is due to enhanced power in the photosphere 
at periodicities near the cut-off.  Possible causes for shocks not accounted by this mechanism 
could be the additional contribution from high-frequency waves when waves near the acoustic cutoff 
frequency are weak (CS97), as well as some instances of non-vertical propagation.

\section{Summary and Discussion}\label{s_discussion}

The temporal sequences of internetwork \CaII 854.2 nm spectra show compelling evidence for abundant acoustic shocks in the mid-chromosphere (Sect. \ref{s_evid_shocks}).  
The evolution of spectral characteristics of the \CaII 854.2 line during these events is in 
complete analogy with 
the more famous \CaII K$_{2V}$ grains, clearly explained by CS97 as due to weak acoustic 
shocks in the mid-chromosphere. 
We have set out to provide a comprehensive view of the occurrence of these shocks, by 
combining a large array of complementary diagnostics, including spectral
information in both photospheric and chromospheric lines; high spatial resolution over an 
extended field-of-view; simultaneous high-resolution magnetic maps from MDI; and observations of two 
different quiet Sun targets with different magnetic topologies, for a period of about 1 hr 
each, at high cadence. We surmise that such a dataset is rather unique, and indeed provided 
much insight into the issue.

\medskip
\noindent$\mathsf{Shock~ identification.}$ 
We have identified the spatio-temporal locations of chromospheric shocks by using 
a velocity criterion on the  \CaII 854.2 nm line, namely the abrupt displacement of the profile from red to 
blue.
This approach provides several intrinsic advantages over the more commonly used intensity 
threshold methods \citep[see e.g.][and references therein]{tritschler_07}:
Using a velocity determination avoids confusion between brightenings from shocks and those 
due to transient small-scale magnetic structures within the internetwork. The latter correspond 
to what has been termed ``magnetic grains'' (RU91) or ``persistent flashers'' \citep
{brandt_92,dwi_08}, and, while as bright as the internetwork grains, they are characterized by 
longer evolutionary timescales. Analysis of filtergrams in \CaII H and K may confuse the intensity
signature from these two different processes. Our shock identification technique also identifies 
the start time of each shock, assuring that  they are counted 
only once during the temporal sequence, something that can be difficult with threshold methods.
Finally, by using the full spectral profile, we reduce the influence of spatially and spectrally
scattered light and photometric errors. The velocity method for identifying shock relies on a 
threshold for the amplitude of the red-to-blue Doppler shift, but while some of the shock 
parameters  did show a dependence on the threshold value (in particular the size, and 
total number), others, such as the temporal characteristics and spatial distribution, did not. 

\medskip
\noindent$\mathsf{Shocks'~ properties.}$
Many of the properties that we derive for the \CaII 854.2 shocks are analogous to those 
reported in the literature for K$_{2V}$ grains,  confirming that the two phenomena 
represent the same physical process. In particular, we find that the \CaII 854.2 shocks often 
appear in bursts, with the typical  interval between shocks
ranging from $\sim$2 to 4 minutes (peaking around 150 s), and intervals between bursts of 5 
minutes and longer.  This intermittent behavior is a reflection of their essentially stochastic 
nature (CS97). No obvious periodicity below 120 s is observed. 

The shocks are not ubiquitous, neither in space nor time. Only  25 to 30\% of the quiet 
internetwork displays five or more shocks during the $\sim$ 1 hour sequences. 
Assuming that shocks  give rise to K$_{2V}$ emission for about 60--80 s for each 
event \citep[e.g.][]{beck_08}, this 
rate of occurrence translates into a measurable K$_{2V}$ emission for about 6--10\% of the 
total spatio-temporal sample, in agreement
with earlier works \citep[e.g.][]{vonuexkull_95,steffens_96}.
Recently, using high resolution \CaII H spectra acquired with POLIS,  \citet{beck_08} 
claimed that in quiet Sun regions
the core of the \CaII H resonance line spends about half of the time in emission. This is  
probably misleading, as they implicitly assumed that 
shocks occur everywhere and continuously with the typical 3 minutes periodicity, an assumption 
obviously not fulfilled on the Sun.

The average shock size of 0.7''--1.0'' arcsec measured from the \CaII 854.2 data is sensibly
smaller than the 1''--2'' arcsec typically reported  for K$_{2V}$ grains. We
believe our measurement comes closer to the true size of the shocks  due to the higher 
spatial resolution reached with IBIS with respect to older works, the limited influence of 
instrumental scattered light on our measurements, and the reduced scattering 
in the solar atmosphere for the \CaII 854.2 line with respect to the H and K. 
A small extension of the shocks explains why earlier spectral observations of the \CaII 
854.2 nm in the quiet internetwork, obtained with a 0.6" pixel size, {did not} display the clear 
sawtooth behavior such as that seen in  Fig. 
\ref{specline} \citep[cf. Fig. 3 in][]{deubner_90}.
The smaller effective size of the \CaII 854.2 shocks, together with the overall better resolution,  
accounts for the larger occurrence rate that we measure in a typical supergranular cell 
($>$20 in our case, vs. 10-20 in the ``best spectroheliograms'', cf. RU91).

\begin{figure*}
\centerline{
\includegraphics[scale=1.0]{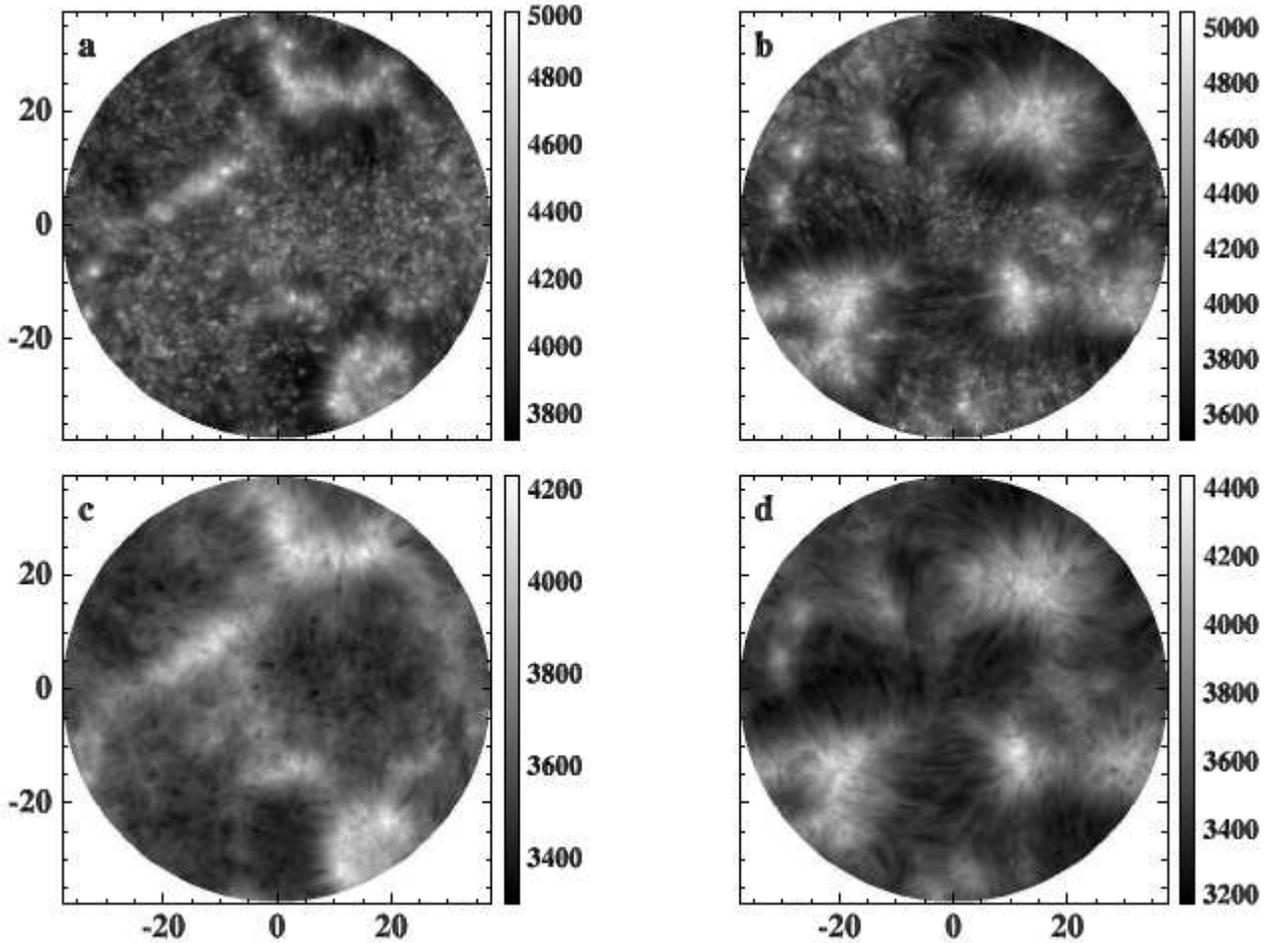}
}
\caption{Maximum (top row) and minimum (bottom row) intensity of the \CaII line core 
throughout the 55 minute sequences. Left: data set 1; right: data set 2. The scale has been
set in radiation temperature (K, see text for details). This representation of 
the data immediately provides a distinction among the three major chromospheric components: 
the bright network; the
acoustic chromosphere where acoustic shocks show up as tiny bright points; and the magnetic 
chromosphere threaded by slanted fibrils that can appear as either dark or bright.}
\label{fig_three_chr}
\end{figure*}

\medskip
\noindent$\mathsf{Spatial ~distribution~and~magnetic~shadows.}$
An important result of our analysis is the extreme inhomogeneity of the spatial distribution of the 
shocks  (Sect. \ref{ss_spatial}). 
It has long been known that shocks can be relatively rare and widely dispersed in the 
supergranular cells \citep[e.g.][]{lites_94}. Indeed we find that up to 50--60\% of the FOV 
display zero or very few shocks, even 
for the case of data set 1 which, being situated at the edge of a coronal hole, arguably has one {of}
the weakest possible magnetic configuration.
However, with our data we clearly show 
that the shocks concentrate in the internetwork  but away from the regions where
 chromospheric fibrils are present. The local magnetic topology outlined by the fibrils is
 very different for the two data sets,  with many short loops closing back from the network 
elements to the near internetwork regions in set 1 and longer and more stable fibrils nearly spanning 
the supergranular cell in set 2 (Fig. \ref{field_line}). 
This nature of the canopy and the spatial distribution of the shocks is not easily predicted 
using only photospheric indicators such as the granulation in the continuum, or the 
location of the network magnetic elements.

A second interesting result is that the total amplitude of the shocks over the course of the 
observations  is directly related to the magnitude of the Fourier power of chromospheric velocities 
with periods
around 3 minutes  (Fig. \ref{f_scatter}). Thus, the shocks represent an important, if not dominant, contribution to the ``3-
minute chromospheric oscillations''
 \citep[e.g.][]{orrall_66,noyes_67},  even if they do occur with an intermittent character. 
Because the shocks are related to large excursions of the 
spectral line, corresponding to several km s$^{-1}$ when parametrized in terms of line core 
Doppler shifts, they can contribute a large fraction of the velocity signal around the acoustic 
cutoff frequency  in classical Fourier analysis, even if they do not occur continuously. The two 
quantities are essentially equivalent, at least in the quiet Sun.
 
These two results fit in nicely with several pieces of evidence reported in the literature about the 
so-called ``magnetic shadows'', and allow us to form a coherent explanation of their existence. First 
introduced by  \citet{judge_01}, the shadows are regions of reduced Fourier oscillatory power at the 
3-minute periodicity, often observed in low- and mid-chromospheric signatures around  (but not 
coincident with) magnetic network elements. They have been reported in intensity observations 
of the continuum at 119 nm obtained with SUMER  \citep{judge_01},  and of the continuum at 
160 nm obtained with TRACE \citep{kri_01}, as well as
in Fourier analysis of  chromospheric velocities by V07 and \citet{lites_94}.

All these works invoke the presence of horizontal magnetic field lines (the ``canopy'') to somehow 
disturb the propagation of normal acoustic waves towards the outer solar atmosphere, for 
example by means
of wave scattering or mode conversion \citep{judge_01,mcintoshjud_01,mcintosh_03}. With our 
data, we prove that the shadows correspond to a strong suppression of the number of acoustic 
shocks.
Further, we show that the shadows are tied to the presence of chromospheric fibrils, i.e. to 
structures with a very distinct physical structure with respect to the surrounding  atmosphere. 
Indeed, we believe they are best compared to the H$\alpha$ mottles and fibrils observed 
around quiet Sun magnetic network, that can be explained either by elevated  structures such 
as spicules, or ``embedded"  ones \citep[see e.g. the discussion in][]{al_04}.
Hence, the upward propagating acoustic waves may be disturbed not so much by
the horizontal field itself, but by the different stratification of the atmosphere.
This should be taken into account in the discussions 
in terms of high- or low- plasma $\beta$ (the ratio of kinetic to magnetic pressure) as the main 
discriminant between an `acoustically-dominated' and a `magnetically-dominated' atmosphere. 
An estimate of $\beta$ that uses the classical, semi-empirical temperature and density values 
(e.g. from VAL) might not be appropriate in these regions.
In the case of elevated structures, it is also possible
that the acoustic shocks develop normally in the low chromosphere whenever photospheric 
conditions are favorable, and the higher-lying  fibrils simply act as a mask, hiding the lower 
layers from view. 
Some support for this scenario can be gathered in  \citet{dewijn_07}
as well as from Movie 2 in Paper I, where the lateral motions of the fibrils at times ``uncover'' 
the presence of bright grains, especially towards the end of their horizontal extension (see also 
Fig. \ref{fig_three_chr}).

\medskip
\noindent$\mathsf{Photospheric ~drivers.}$
The analysis of Sec. \ref{ss_phase} clearly {shows} that the \CaII 854.2  shocks partake in the 
general chromospheric dynamics, responding to acoustic waves propagating from lower layers. 
The phase relationships between photospheric and chromospheric velocities in regions 
undergoing shocks are 
fully consistent with earlier results obtained for general internetwork areas, proving the 
propagation of acoustic waves up to 10--12 mHz \citep{deubner_90}. The occurrence of shocks 
in this general oscillation pattern must then be triggered by complementary circumstances. By 
means of a wavelet analysis we find that instances of high photospheric velocity power at 
periodicities between 120 and 200 s (5--8 mHz) are well correlated with the occurrence of 
chromospheric shocks. Tellingly, the correlation is maximized when allowing for a delay of 
$\sim$120 s between the photospheric and chromospheric signal, indicating a photospheric 
perturbations propagating upward at the speed of sound over a $\sim$ 700--1000 km height 
difference. 

The small dimension of the shocks is not completely correlated with the piston's size -- the 
analysis of Sect. \ref{ss_wavelets} shows how areas of coherent photospheric oscillations at the 
relevant periodicity have lateral sizes of up to several arcsec.
However, it is important to note that these areas completely encircle the {majority of the} pixels that develop 
shocks (left panels in Fig. \ref{fig_wavelet_maps}). 
All of these results are consistent with the analysis of  CS97 that state  how the appearance of 
the 
bright grains, including their size, is fully determined  by powerful waves near 
or just above the acoustic cutoff frequency interfering with 
higher frequency waves. The interference pattern that determines the actual size and 
recurrence of shocks will be contained within, but not necessarily coincident with,  the regions 
of enhanced photospheric velocity at the dominant frequencies.

The strength of photospheric motions is not the only factor shaping  the ``quiet'' 
chromosphere and its dynamics. We find that
the presence of magnetic structures has a profound influence on what occurs in the lower 
chromosphere and, in particular, whether the acoustic mechanism outlined above can operate 
undisturbed. Many areas with suitable photospheric 
dynamics do not develop shocks (the ``shadows'') and show a decreased coherence 
between the vertical photospheric and chromospheric velocity signals at all frequencies 
(panels {\it c} and {\it d}  of Fig. \ref{phase}).
The dynamics of the fibrils present in the shadows' areas probably is most-closely related to the oscillatory behavior 
at their footpoints than to the directly
underlaying layers.
 Interestingly, we note that pixels developing a small number 
of shocks have a lower coherence 
than the ones developing many shocks; we assume that this is related to a 
mechanism similar to the shadows, just operating on smaller scales. Several hints support this 
hypothesis, such as:  the diffuse appearance of the \CaII line core images (Fig. \ref{fig_fov}, especially 
panel {\it d}) that might betray short, unresolved fibrils; the correlation between the presence of 
small, transient magnetic structures observed in the MDI HR maps and the reduction in total 
shocks numbers (Fig. \ref{f_mdi_shocks}); the  lack of chromospheric shocks in areas 
corresponding to enhanced photospheric velocity power patches, if there exist nearby small 
scale magnetic structures (Sect. \ref{ss_wavelets}).

\medskip
\noindent$\mathsf{Vertical~ propagation.}$
Throughout this paper, we searched for correlations between photospheric and chromospheric 
signal under the assumption of vertical propagation. This is of course a limitation of our analysis and 
has been invoked often as one of the major problems in the CS97 study of K$_{2V}$ grains 
\citep[e.g.][]{ulmschneider_05}.
However, there are several indications that this assumption is not misguided. First,
the level of coherence between the photospheric and chromospheric velocities at the dominant 
frequency of $\sim$5 mHz is very high, especially for the quieter data set 1 (Fig. \ref{phase}). 
This would not be the case if the propagation {were not} predominantly vertical. Indeed,
we find a much lower coherence value
 for the case of the 
chromospheric fibrils, which have an obviously slanted geometry. 
Second, the wavelet analysis of Sect. \ref
{ss_wavelets}  shows how patches of high photospheric power at the relevant periodicity well 
coincide with the appearance of shocks in the chromospheric areas directly above. As the 
photospheric patches of high power extend for 1500--3000 km in the horizontal dimension, i.e. 
at least twice as much as the vertical separation between the regions of formation of the \FeI 
and \CaII lines, a small deviation from the vertical (say below 30$^\circ$) would also 
not cause any drastic difference in our results.

The wavelet analysis shows that in some instances chromospheric shocks are measured even 
in absence of high photospheric power; these could be obvious candidates for a search on the 
effects of inclined propagation, and deserve further investigation. They could also be related to 
the presence of high frequency waves or photospheric power just below our cutoff for enhanced power. 
Still, it appears that small magnetic structures 
represent a more important factor in the shaping of the 3-D chromospheric dynamics, reducing 
the overall number of shocks by large amounts (Sect. \ref{ss_wavelets}).

\medskip
\noindent$\mathsf{Three~quiet~chromospheres.}$
From our results, we propose that the ``quiet'' lower-chromosphere should be considered to 
comprise at least three components, as somewhat anticipated in \citet{judge_01} 
and remarked in Paper I. {These three components are:} 1- the 
purely acoustic chromosphere, where the strong photospheric motions at frequencies near the 
acoustic cut-off can propagate upwards undisturbed and develop into shocks; 2- the network 
and internetwork magnetic elements, that open up partially from the photosphere and are 
heated by a mechanism still unexplained \citep[see][and references therein]{hasan_08};
and 3- the  inclined fibrils that magnetically connect horizontally separated areas. The fibrils can 
span as long as a supergranular cell (as in data set 2) or as short as few arcseconds (as in the 
smallest shadows in Fig. \ref{map1} {\it d}); in any case they define chromospheric volumes 
with very different dynamical properties than the other two components. In particular, they show a reduced 
chromospheric {\it vertical} velocity power at all frequencies (see also Paper I); distinctly longer evolutionary 
timescales (V07); a lack of vertical coherence between the photospheric and chromospheric 
signals (Sect. \ref{ss_phase}).  

These three components can be immediately identified by using maps of the \CaII 854.2 line 
core intensity, a reliable indicator of temperature (Cauzzi {\it et al.}, in preparation). In particular, we 
show in Fig. \ref{fig_three_chr} the maximum (top row) and minimum (bottom row) value of the 
line core intensity, measured for each spatial pixel within the 55 min observing sequences. In the 
Figure the intensity has been converted to radiation temperature by scaling the average 
intensity computed over the full spatio-temporal sample to the calibrated atlas intensity of \citet
{neckel_99}
and applying the inverse Planck function (the average intensity corresponds to 3800 K).
Obviously this definition does not represent the real kinetic temperatures as 
the line is not formed in LTE, but it does provide an indication for the magnitude of the variations.
In Fig. \ref{fig_three_chr} the acoustic shocks stand out clearly  in the maximum intensity maps 
as bright specks with a quasi-granular appearance (but with smaller dimensions) and 
temperatures  up to 4200 K. 
At times, especially in data set 2, the shocks can even be discerned in amongst the long fibrils, supporting 
the idea of the latter operating as a switch in between the lower chromosphere and the 
observer. The canopy-free internetwork regions where there are  significant numbers of shocks are also the regions of the 
lowest intensity, and presumably lowest temperature, in the entire sequence, with 
radiation temperature values as low as 3400 K.
The magnetic elements appear also as distinct point-like sources, but larger and brighter than 
the acoustic shocks, with 
temperature excursions ranging from a minimum of 4000 to maxima of 4700 K.
Finally, the slanted fibrils have a very different morphology and might be better defined as the 
``diffuse'' sources in both the maximum and minimum intensity maps. Two things about them 
are particularly remarkable: they are the component whose intensity (temperature) changes the 
least between maximum and minimum values,
with largest values close to the network points and {cooler} ends towards the internetwork; and 
they cover a large fraction of the FOV, even for data set 1. The magnetic influence appears 
very pervasive over the quiet Sun.

Finally, it would appear that Fig. \ref{fig_three_chr} also offers a definitive statement on 
the issue of the spatial correspondence between small scale magnetic structures and K$_{2V}$ 
bright points \citep[e.g.][]{sivaraman_00}. Indeed, many such small magnetic concentrations  
are visible in the internetwork portion of the FOV in both of our data sets, and they do 
correspond to enhanced bright ``points'', some of them only 1--2'' wide. However, they also are 
clearly distinguished from the acoustic grains by the  diffuse cocoon around them, 
made up 
most probably of unresolved fibrils. They appear just like miniature versions of normal network 
elements \citep{kri_01}. The analysis of Sect. \ref{s_mag} makes it clear that these bright 
structures do not participate in the typical dynamics of acoustic shocks, making them K$_{2V}$ 
bright ``points'', but not K$_{2V}$ bright grains!
Claims to the contrary, such as those of \citet{sivaraman_00}, stem from the limited coverage of the 
available datasets or the limited number of co-spatial magnetic 
and K$_{2V}$ measurements.

\section{ Conclusions}\label{s_concl}

In our analysis, we reach two main conclusions, both of which are represented in 
Fig. \ref{fig_three_chr}. 

The first is that temporally resolved spectra of the \CaII 854.2 nm 
provide a comprehensive picture of the occurrence of acoustic shocks in the quiet solar atmosphere. 
We determine that these shocks are 
a fundamental
part of the general chromospheric dynamics, as they
essentially make up the bulk of the three-minute oscillations in 
the 
middle chromosphere.
We find that the shocks  occur in direct response (after a 120 second delay) to 
powerful
photospheric oscillations 
near the acoustic cutoff frequency, with a primarily vertical propagation. No major role is found for 
waves with periods below 120 seconds or with a slanted propagation, consistent with the 1-D purely hydrodynamical 
simulations of \citet{carlsson_97}. Portions of the quiet chromosphere are then the site of 
brief episodes when the acoustic shocks dramatically enhance the temperature (Fig. \ref{fig_three_chr} {\it a} and {\it b})  
mixed with extended periods of lower temperatures  (Fig. \ref{fig_three_chr} {\it c} and {\it d}), as opposed to
the steady-state mean temperature rise 
typically assumed in semi-empirical models 
\citep[see the discussion in][]{carlsson_95,carlsson_07}.

The second, more important conclusion is 
that
these acoustic processes are very often 
significantly disturbed by the presence of magnetic field, even in the quiet Sun. 
Both at the network and (surprisingly) at internetwork scales, the signature of 
acoustic shocks is suppressed both in and around the footpoints of magnetic structures. Hence the 
distribution of shocks is considerably reduced compared to what would be inferred from 
the amplitude of the photospheric oscillations alone. 
The main player in this respect appears to be the magnetic 
topology,  and in particular  whether the field has an open or closed configuration. The areas of 
modified acoustic behavior coincide with the presence of chromospheric fibrils, betraying an 
atmosphere that is stratified differently with respect to the classical 1-D view. 

These results provide an important hint concerning the longstanding controversy 
about why
the dynamical model of CS97 fails to predict the ubiquitous emission observed in UV lines \citep
[e.g.][]{kalkofen_99,kalkofen_01,carlsson_07,avrett_08}. 
Our analysis is based on observations in the \CaII 854.2 nm line, a spectral signature formed 
relatively low in the atmosphere. Our data provide uncontroversial evidence that even at these 
heights  the quiet atmosphere is highly structured by pervasive magnetic 
fields that connect different locations.
The  presence of fibrils is a 
clear testimony of these inter-connections\footnote{We note that this magnetic 
structuring  has been observed for decades in H$\alpha$, a diagnostics that has 
figured little in the discussion on chromospheric heating.}. It is then natural to assume that 
other spectral diagnostics forming in higher, less dense parts of the atmosphere,  experience 
such a magnetic structuring to an even larger extent, much as advocated by \citet{ayres_02}. 

Far UV lines and continua are one such diagnostic, and they figure prominently in the 
derivation of the temperature structure in semi-empirical models, often as spatial and temporal 
averages \citep[e.g.][]{avrett_08}. It follows that a substantial fraction of the UV flux  
could originate in magnetic structures, even in the {\it quiet} internetwork chromosphere. This 
could happen either because of the volume filling of the magnetic field at these heights,  or
because of the dominant weight of magnetic-related brightness in the averages: Fig. \ref
{fig_three_chr} illustrates how magnetic concentrations introduce a brightness component 
with a different spatial distribution and temporal evolution than for the purely acoustic case. This 
happens on spatial scales at the limit of the resolution of many modern instruments, including 
SUMER.  Any averaged or marginally resolved observation of the chromospheric flux might then 
contain a sizable magnetic component, in ways not immediately predictable by using only 
photospheric indicators. 
A dominance of the magnetic topology in the quiet upper chromosphere
would also explain why UV observations obtained with SUMER have produced so many 
disparate results. These range from the clear signature of upward propagating acoustic waves 
reported by \citet{wikstol_00}
all the way from the photosphere up to the base of the corona, as could happen in the case of 
a coronal hole, to instances of complete disconnection between the observed lower and upper 
chromospheric dynamics \citep{judge_01,judge_03}, as would be the case for regions with a 
locally closed magnetic topology.  
In this scenario, obviously there is no need to reconcile the predictions of CS97, a purely 
non-magnetic model, with the UV observations, as they pertain to completely different physical 
regimes.

We thus conclude that the  radiative losses in the ``non-magnetic'' (quotes are necessary at this point) 
solar chromosphere are strongly influenced, and possibly dominated, by processes related 
to magnetic fields.
We remark that this same conclusion has been reached by Judge and collaborators 
\citep{judge_98,judge_03,judge_04}, in a precise analysis of UV emission lines  for both the 
case of the Sun and other stars with convective envelopes. 
The presence of fibrils in large fractions of the quiet atmosphere provides a plausible 
agent for such dominance. On the one hand, they disrupt the normal vertical propagation 
of acoustic waves from lower layers, strongly reducing the effects of a purely acoustic 
mechanism in shaping the chromosphere. On the other, they provide the means to 
transport energy of magnetic origin, or mediated by the magnetic elements, from the network 
footpoints towards the center of the supergranular cells.
We note that this idea is consistent  with several recent works that
have called into question whether the photospheric acoustic flux at high frequencies is sufficient 
to compensate the chromospheric radiative losses \citep
{fossum_05,fossum_06} or, even more pertinent,  whether different energy sources 
 such as {\it low-} frequencies magneto-acoustic waves, or atmospheric gravity waves,
might play a more important 
role  \citep{jefferies_06,straus_08}.

\acknowledgements{This work was partially supported by PRIN-INAF 2007: ``Scientific exploitation
of the Interferometric Bidimensional Spectrometer (IBIS).
Magnetic structuring of the lower solar atmosphere.''\\
The authors are grateful to the DST observers D. Gilliam, M. Bradford and J. Elrod, whose
patience and skills are greatly appreciated. 
IBIS was
built by INAF--Osservatorio Astrofisico di Arcetri with contributions 
from the Universit\`a di Firenze and the Universit\`a di
Roma ÒTor VergataÓ. Further support for  
IBIS operation was provided by the Italian MIUR and MAE,
as well as NSO. NSO is operated by the Association of Universities
for Research in Astronomy, Inc. (AURA), under cooperative
agreement with the National Science Foundation. SOHO is a
project of international cooperation between ESA and NASA. Wavelet software was provided by 
C. Torrence and G. Compo, and is available at \url{http://atoc.colorado.edu/research/
wavelets/}. We made much use of the NASA's Astrophyics Data System.}



\begin{appendix}

\section{Identification of shocks}\label{app_shocks}

\subsection{The velocity threshold approach}\label{ss_vel}

A way to  identify acoustic shocks ``events''  is through the sawtooth peaks in the 
velocity field, defined via the Doppler shift of the line intensity minimum. The small white 
crosses in Fig. \ref{specline} panel {\it a} show this parameter for the particular pixel. As 
apparent from both Fig. \ref{specline} and \ref{fig_plotshock}, during the upward phases of the 
shock the spectral line broadens considerably in the blue wing, probably leading to an 
underestimate of the actual value of the vertical velocity by using the simple Doppler shift of the 
minimum. However, such an effect does not seem particularly severe (compare for example  
Fig. 7 in Pietarila et al. 2006), and  the sign of the recovered velocity is always correct.

Following the temporal behavior of the velocity, for each spatial pixel a shock event is then 
defined to occur at the time when the velocity experiences an abrupt jump from large  positive 
to large negative values (downward to upward motion). As a temporal threshold for this velocity 
jump, we utilize a maximum interval of two time steps, i.e. a 38 s window. 
The threshold for the amplitude of the velocity jump is set at a value of  4 km s$^{-1}$. This is 
safely above the r.m.s value of  $\sim$1 km s$^{-1}$ measured for chromospheric velocities at 
the dominant periodicity, and allows us to identify the most energetic events.  Experiments with 
lower and higher values of the velocity threshold cause a different number of shock events to 
be counted, but  do not alter the main conclusions about their  overall spatio/temporal 
distribution.

With these two combined criteria we can identify the presence of an acoustic shock at any given 
time or location within the FOV, and hence derive a temporal series of binary maps defining 
their spatial distribution, similar to the examples displayed in Fig. \ref{map1} {\it a} and {\it e}. A 
value of the shock strength can also be defined as the maximum velocity difference between 
downward and upward velocities during the shock. Given the limitations of the Doppler shift of 
the intensity minimum, described above, both of these values can be considered as lower limits 
to the actual distributions.

\subsection{The Proper Orthogonal Decomposition approach}\label{ss_pod}

As an independent validation of the results obtained through the velocity threshold method, we 
also attempted a more general approach, based  on an analysis of the whole spectral shapes. 
This approach makes use of the  Proper Orthogonal Decomposition (POD) technique, initially 
developed to study coherent structures in channel flow turbulence and recently applied to 
various astrophysical problems \citep{rees_00,carbone_02,ruzm_04,vecchio1,vecchio2}. As the 
POD technique is not yet widely known, we will describe it in some detail in the following.

\subsubsection{The POD method}

In the framework of the POD, a spatio-temporal field $u(r,t)$ is decomposed as $\sum_{j=0}^
{\infty}a_j(t)\Psi_j(r)$.
The orthonormal basis functions, $\Psi_j$, are not given a priori but obtained from the 
experiment, and can hence assume the proper functional shape
of the phenomenon. The time coefficients $a_{j}(t)$, computed from the projection of the data 
on the corresponding basis functions, represent the time evolution of the $j$-th mode 
associated with that eigenfunction.
The modes are then ordered according to a parameter, the ``energy'' content of fluctuations 
associated with them, quantifying the relative contribution of each mode to the signal 
reconstruction. The most energetical mode ($j=0$) is associated to the average contribution to 
the signal. For more details, we refer to  the text by \citet{lum_96}. 

\subsubsection{Application of POD to \CaII 854.2 profiles}

We apply the POD to analize the temporal evolution of the spectral profiles (like those shown in 
Fig. \ref{specline}) for each pixel of the FOV separately, considering the (wavelength-
dependent) intensity fluctuations around the mean profile. Thus, in our application the the POD 
decomposition of a wavelength-time spectrum will be
$$ s(\lambda,t)=\sum_{j=0}^{\infty}a_j(t)\Psi_j(\lambda) $$
namely, the role of spatial coordinate is played by the wavelength. For most of the pixels, we 
find that the 99\%  of the total energy is contained in the first four
POD modes (including the average, $j=0$), i.e. these are the only significant modes in the 
signal
reconstruction.

\begin{figure*}
\centerline{
\includegraphics[scale=1.0]{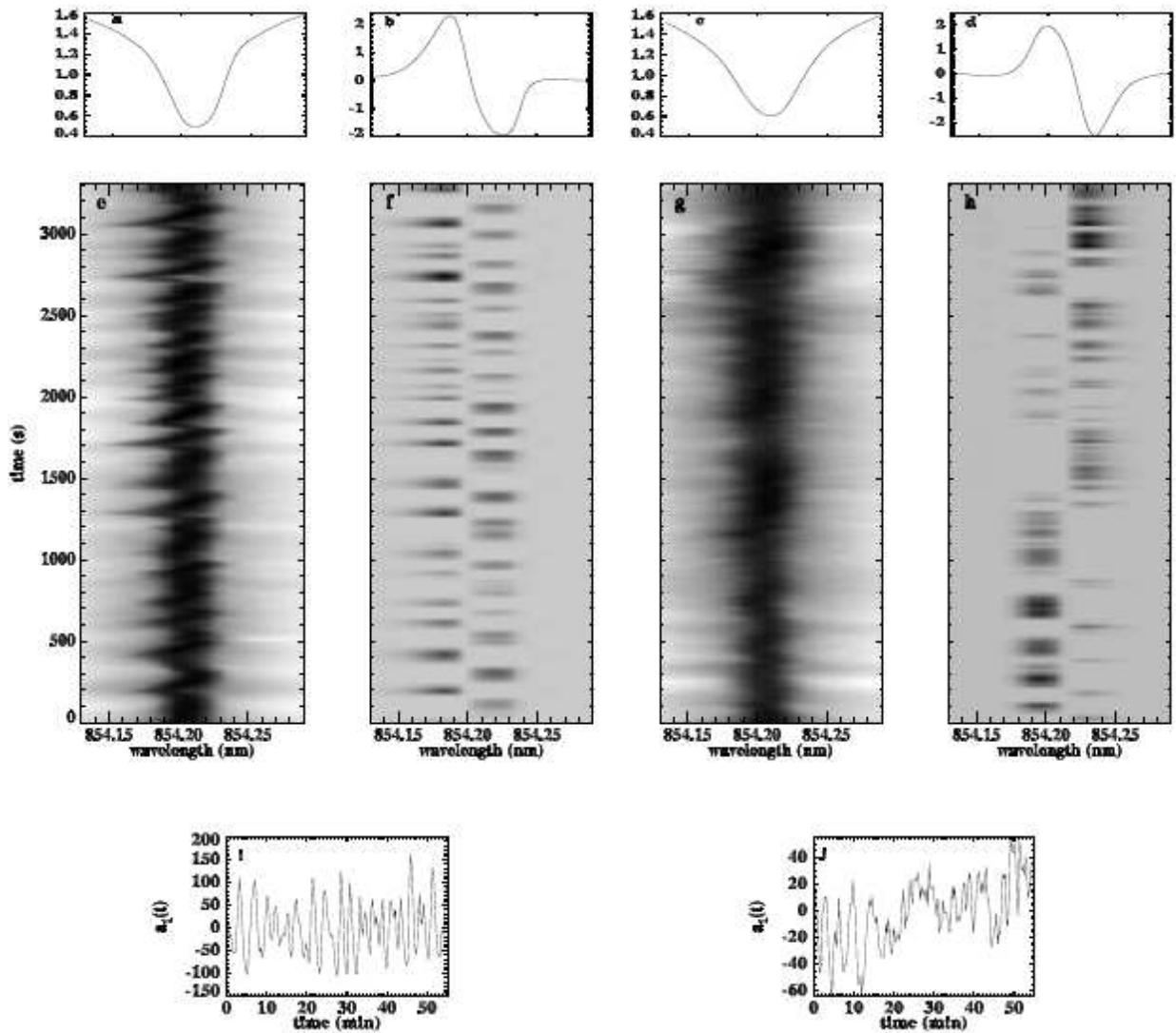}
}
\caption{Example of POD analysis for two pixels from data set 1. Panels {\it a--b, e--f} refer to a 
pixel displaying a large number of shocks with 3 min periodicity. {\it a}: Eigenfunction of the 
mode $j=0$  (average of the signal). {\it b}:  Eigenfunction describing the shocks contribution, 
displaying the typical asymmetric shape. Since the eigenfunction are orthonormal their 
amplitude is arbitrary. {\it e}: Actual measured spectra. {\it f}: Contribution to the profiles in {\it e} 
reconstructed using only the POD mode associated to an eigenfunction like {\it b} (mode 1 in 
this case). For clarity, only the positive portion of the reconstructed function is displayed. Panels 
{\it c--d, g--h}: As before, but for a pixel where only few, energetically unimportant  shocks are 
found by the POD analysis (shock mode = 3).
Panels {\it i--j}: POD temporal coefficients used in the reconstructions of panels {\it f} and {\it h}.  
A 3 minute periodicity is evident in panel {\it i}. 
}
\label{eigf_pod}
\end{figure*}

\begin{figure*}[t]
\centerline{
\includegraphics[scale=1.0]{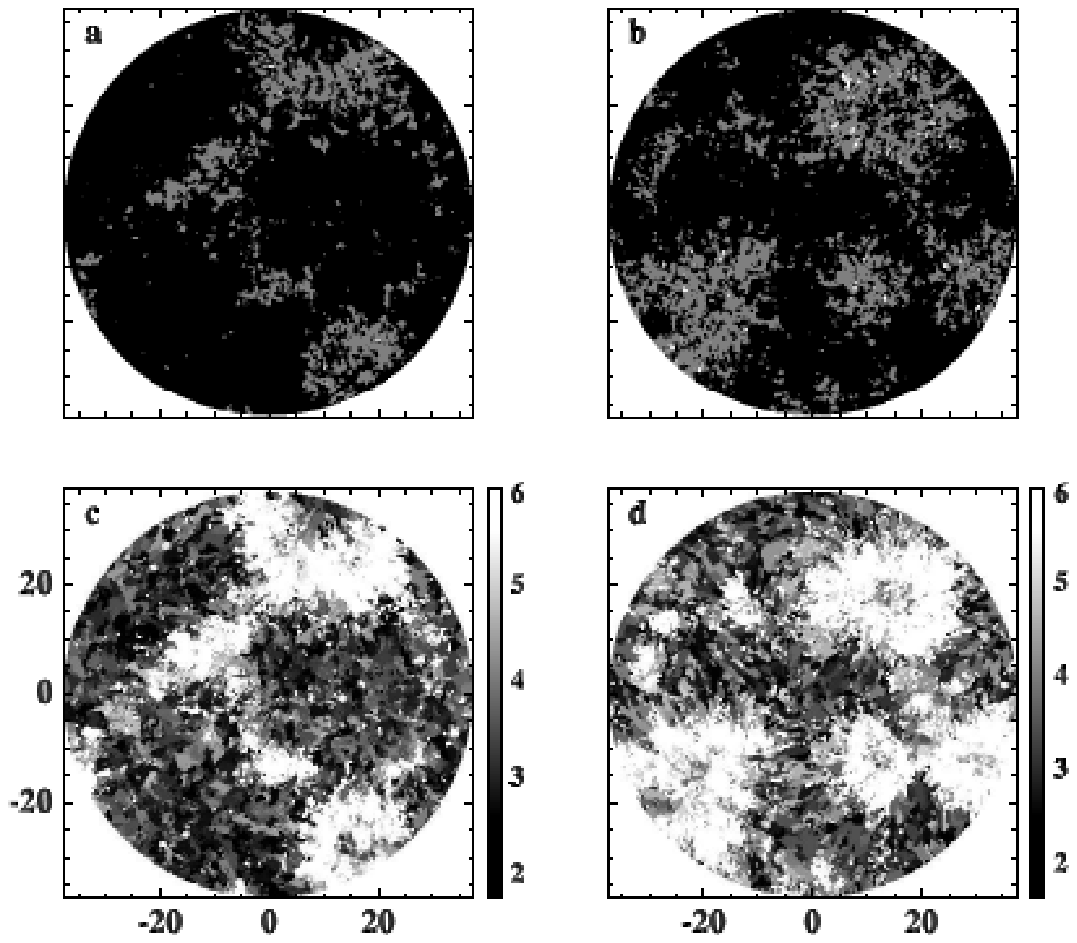}
}
\caption{Panels {\it a} and {\it b}: maps of mode pertaining to the shock eigenfunctions, for data set 1 and 2.
 {\it c} and {\it d}: maps of dominant periodicity in coefficients $a_i(t)$, for data set 1 and 2. 
Spatial axes in arcsec.}
\label{mode_eigf}
\end{figure*}

Panel {\it e} in Fig. \ref{eigf_pod} displays, as an example, the actual temporal evolution of the 
spectrum corresponding to an internetwork pixel in data set 1, where we find evidence for 
several shocks. Panel {\it a} shows the corresponding POD eigenfunction $\Psi_0(\lambda)$ of 
the mode $j=0$, i.e. the temporally averaged spectrum.  
The chromospheric  steepening of acoustic oscillations, which eventually will lead to shock 
development, 
induce  an intensity fluctuation characterized by a well defined behavior, i.e. a large 
displacement of the line from the average position, with an asymmetric contribution in the two 
wings.
The POD analysis isolates this behavior in a single mode, whose eigenfunction has the typical 
shape shown in panel {\it b} of Fig. \ref{eigf_pod}. It is characterized by two opposite lobes of 
different amplitude, indicating positive and negative contribution to the average profile towards 
the blue and the red wavelengths. The sign of the time coefficient (to be multiplied to the 
eigenfunction) selects when the positive contribution is in the red or in the blue part of the 
spectrum.

Thus for each pixel the mode associated to the (possible) shock contribution can be identified 
by looking at the shape of the eigenfunctions, including the lobe asymmetry.  
Suppose that for the spectrum of Fig. \ref{eigf_pod} {\it e} this is the mode $j=i$.
The wavelength-time intensity contribution of this mode to the actual spectral evolution can be 
expressed as $a_i(t)\Psi_i(\lambda)$, and is
shown in panel {\it f} of Fig. \ref{eigf_pod} (for clarity we have displayed only its positive part).  It 
illustrates how the modifications to the spectral profiles induced by this mode are in general 
stronger and more extended in the blue wing of the line, as well as the fact that, on average, 
the variations in the red wing last longer than in the blue wing. The latter property is present 
also in the simulations of CS97.
For this particular pixel, $i=1$, i.e. the mode characterizing the eigenfunction $\Psi_i(\lambda)$ 
is  the one with the highest ``energy'' content of fluctuations. 

\subsubsection{Dominant POD modes and periodicities}\label{ss_pod_period}

Obviously, not all the pixels within the FOV have spectra similar to those displayed in panel {\it 
e} of Fig. \ref{eigf_pod}. Panel {\it g} of the same  Figure provides an illuminating example, 
showing the temporal evolution of the spectrum for a pixel pertaining to a fibril within the FOV of 
data set 1. The average spectrum (POD eigenfunction of the mode  $j=0$) is given in panel {\it 
c}, the ``shock-mode'' eigenfunction for $j=i$ is shown in panel  {\it d}, and the  POD 
reconstruction for this mode is displayed in panel {\it h}. 
For this case,
the mode characterizing the eigenfunction $\Psi_i(\lambda)$ is $i=3$, i.e. with a much lower 
energy than the example of panel {\it e}. 

Panels {\it a} and {\it b} of Fig. \ref{mode_eigf} display the spatial distribution, for both data sets, 
of the mode $i$ defined above. (As said before, modes with $i > 3$ contain an energy less than 
1\% and have not been considered.) For the majority of the internetwork pixels $i=1$, i.e. the 
``shock-mode'' is the most energetic component of the intensity fluctuation. However, there 
exist  large portions of the FOV, apparently pertaining to the photospheric internetwork but 
surrounding quite extensively the magnetic elements, for which the relevant eigenfunction 
corresponds to the mode $i = 2$. Their typical energy fluctuation content is less than half that 
pertaining to the pixels with shock-mode $i=1$. This effect is more prominent in data set 2, in 
correspondence to a larger extension of the fibrillar structures visible in the line core images.

The coefficients $a_j(t)$ provide the temporal behavior of the different POD modes for each 
spatial pixel. For example, the coefficients used in the recostructions of Fig.  \ref{eigf_pod} {\it f} 
and {\it h} are plotted in the same figure  (panels {\it i} and {\it j}), and show that for the first 
case the ``shock'' eigenfunction has a clear periodicity nearing 150 s, while for the second one 
the temporal coefficient has a more erratic behavior. Maps of the dominant periodicities in the 
$a_j(t)$ coefficients, for each pixel and for the shock-associated $j=i$ POD mode, are reported 
in panels {\it c} and {\it d} of Fig. \ref{mode_eigf}. They have been derived from a Fourier 
analysis of the  coefficients, but the same result is obtained if using a wavelet analysis of the 
time series, by considering the  wave period containing the highest number of wave packets 
with significant power.  In both data sets, for most of the pixels characterized by the $i = 1$ 
POD mode, the dominant periodicity is $\sim$ 200 s  and slightly shorter (frequencies between 
5 and 7 mHz).  This is consistent with the typical chromospheric periodicity of $\sim$ 3 minutes 
for acoustic oscillations, deriving from the atmospheric  filtering of the photospheric $p$-modes. 
The association between $i=1$ POD mode and this periodicity essentially reflects the strong 
steepening of the acoustic waves propagating in the chromosphere. We note that the two data sets display 
differences in this association, with  a more  fragmented distribution of the short-periodicity 
areas, and a stronger presence of longer periodicities (4 mHz or less) in data set 2. 

We further note that some spatial locations within the magnetic network areas display ``shock-
associated'' eigenfunction $i=1$  as well, but are characterized by longer periodicities 
(frequency below 4 mHz). They will be considered in more detail in a future paper.

\subsubsection{POD shocks}

Finally, in order to identify the occurrence of shocks from a pattern such as that displayed in 
Fig. \ref{eigf_pod} {\it f} or {\it h},  we can simply count the events characterized by a red peak 
followed by a blue peak, with the jump occurring within the same temporal delay  defined before 
(38 s),
without applying any threshold to the amplitude of the events.  Rather, an amplitude can be 
defined as the intensity difference of the two peaks of the eigenfunction  ($a_i(t)\Psi_i(\lambda)$) on either (temporal) side of the shock. Such difference is related to the variations induced by 
the shock over an average profile for the pixel, and is then again a measure of the strength of 
the event.  
As in the velocity analysis,  repeating this calculation for each pixel we can derive a
temporal series of binary maps, displaying the shocks' spatial distribution, as well as maps of 
their amplitudes. 
Panels {\it a} and {\it e} of Fig. \ref{map1} show typical ``shocks maps'', for a time around the 
middle of the temporal sequence.

\end{appendix}


\begin{thebibliography}{80}
\expandafter\ifx\csname natexlab\endcsname\relax\def\natexlab#1{#1}\fi

\bibitem[{{Al} {et~al.}(2004){Al}, {Bendlin}, {Hirzberger}, {Kneer}, \&
  {Trujillo Bueno}}]{al_04}
{Al}, N., {Bendlin}, C., {Hirzberger}, J., {Kneer}, F., \& {Trujillo Bueno}, J.
  2004, \aap, 418, 1131

\bibitem[{{Anderson} \& {Athay}(1989)}]{anderson_89}
{Anderson}, L.~S. \& {Athay}, R.~G. 1989, \apj, 336, 1089

\bibitem[{{Avrett} \& {Loeser}(2008)}]{avrett_08}
{Avrett}, E.~H. \& {Loeser}, R. 2008, \apjs, 175, 229

\bibitem[{{Ayres}(2002)}]{ayres_02}
{Ayres}, T.~R. 2002, \apj, 575, 1104

\bibitem[{{Beck} {et~al.}(2005){Beck}, {Schmidt}, {Kentischer}, \&
  {Elmore}}]{beck_05}
{Beck}, C., {Schmidt}, W., {Kentischer}, T., \& {Elmore}, D. 2005, \aap, 437,
  1159

\bibitem[{{Beck} {et~al.}(2008){Beck}, {Schmidt}, {Rezaei}, \&
  {Rammacher}}]{beck_08}
{Beck}, C., {Schmidt}, W., {Rezaei}, R., \& {Rammacher}, W. 2008, \aap, 479,
  213

\bibitem[{{Biermann}(1948)}]{biermann_48}
{Biermann}, L. 1948, Zeitschrift fur Astrophysik, 25, 161

\bibitem[{{Brandt} {et~al.}(1992){Brandt}, {Rutten}, {Shine}, \&
  {Trujillobueno}}]{brandt_92}
{Brandt}, P.~N., {Rutten}, R.~J., {Shine}, R.~A., \& {Trujillobueno}, J. 1992,
  in Astronomical Society of the Pacific Conference Series, Vol.~26, Cool
  Stars, Stellar Systems, and the Sun, ed. M.~S. {Giampapa} \& J.~A.
  {Bookbinder}, 161--+

\bibitem[{{Carbone} {et~al.}(2002){Carbone}, {Lepreti}, {Primavera},
  {Pietropaolo}, {Berrilli}, {Consolini}, {Alfonsi}, {Bavassano}, {Bruno},
  {Vecchio}, \& {Veltri}}]{carbone_02}
{Carbone}, V., {Lepreti}, F., {Primavera}, L., {et~al.} 2002, \aap, 381, 265

\bibitem[{{Carlsson}(2007)}]{carlsson_07}
{Carlsson}, M. 2007, in Astronomical Society of the Pacific Conference Series,
  Vol. 368, The Physics of Chromospheric Plasmas, ed. P.~{Heinzel},
  I.~{Dorotovi{\v c}}, \& R.~J. {Rutten}, 49--+

\bibitem[{{Carlsson} {et~al.}(2007){Carlsson}, {Hansteen}, {de Pontieu},
  {McIntosh}, {Tarbell}, {Shine}, {Tsuneta}, {Katsukawa}, {Ichimoto},
  {Suematsu}, {Shimizu}, \& {Nagata}}]{carlssonetal_07}
{Carlsson}, M., {Hansteen}, V.~H., {de Pontieu}, B., {et~al.} 2007, \pasj, 59,
  663

\bibitem[{{Carlsson} {et~al.}(1997){Carlsson}, {Judge}, \&
  {Wilhelm}}]{carlssonetal_97}
{Carlsson}, M., {Judge}, P.~G., \& {Wilhelm}, K. 1997, \apjl, 486, L63+

\bibitem[{{Carlsson} \& {Stein}(1995)}]{carlsson_95}
{Carlsson}, M. \& {Stein}, R.~F. 1995, \apjl, 440, L29

\bibitem[{{Carlsson} \& {Stein}(1997)}]{carlsson_97}
{Carlsson}, M. \& {Stein}, R.~F. 1997, \apj, 481, 500

\bibitem[{{Carlsson} \& {Stein}(2002)}]{carlsson_02}
{Carlsson}, M. \& {Stein}, R.~F. 2002, in ESA SP-505: SOLMAG 2002. Proceedings
  of the Magnetic Coupling of the Solar Atmosphere Euroconference, ed.
  H.~{Sawaya-Lacoste}, 293--300

\bibitem[{{Cauzzi} {et~al.}(2008){Cauzzi}, {Reardon}, {Uitenbroek},
  {Cavallini}, {Falchi}, {Falciani}, {Janssen}, {Rimmele}, {Vecchio}, \&
  {W{\"o}ger}}]{cauzzi_08}
{Cauzzi}, G., {Reardon}, K.~P., {Uitenbroek}, H., {et~al.} 2008, \aap, 480, 515

\bibitem[{{Cauzzi} {et~al.}(2007){Cauzzi}, {Reardon}, {Vecchio}, {Janssen}, \&
  {Rimmele}}]{cauzzi_07}
{Cauzzi}, G., {Reardon}, K.~P., {Vecchio}, A., {Janssen}, K., \& {Rimmele}, T.
  2007, in ASP Conference series, Vol. 368, The physics of chromospheric
  plasmas, ed. P.~{Heinzel}, I.~{Dorotovi\v{c}}, \& R.~R. J.

\bibitem[{{Cavallini}(2006)}]{cavallini_06}
{Cavallini}, F. 2006, \solphys, 236, 415

\bibitem[{{Cram} {et~al.}(1977){Cram}, {Brown}, \& {Beckers}}]{cram_77}
{Cram}, L.~E., {Brown}, D.~R., \& {Beckers}, J.~M. 1977, \aap, 57, 211

\bibitem[{{Cram} \& {Dame}(1983)}]{cram_83}
{Cram}, L.~E. \& {Dame}, L. 1983, \apj, 272, 355

\bibitem[{{Cuntz} {et~al.}(2007){Cuntz}, {Rammacher}, \& {Musielak}}]{cuntz_07}
{Cuntz}, M., {Rammacher}, W., \& {Musielak}, Z.~E. 2007, \apjl, 657, L57

\bibitem[{{de Wijn} {et~al.}(2007){de Wijn}, {De Pontieu}, \&
  {Rutten}}]{dewijn_07}
{de Wijn}, A.~G., {De Pontieu}, B., \& {Rutten}, R.~J. 2007, \apj, 654, 1128

\bibitem[{{de Wijn} {et~al.}(2008){de Wijn}, {Lites}, {Berger}, {Frank},
  {Tarbell}, \& {Ishikawa}}]{dwi_08}
{de Wijn}, A.~G., {Lites}, B.~W., {Berger}, T.~E., {et~al.} 2008, \apj, 684,
  1469

\bibitem[{{Deubner} \& {Fleck}(1990)}]{deubner_90}
{Deubner}, F.-L. \& {Fleck}, B. 1990, \aap, 228, 506

\bibitem[{{Fontenla} {et~al.}(2007){Fontenla}, {Balasubramaniam}, \&
  {Harder}}]{fontenla07}
{Fontenla}, J.~M., {Balasubramaniam}, K.~S., \& {Harder}, J. 2007, \apj, 667,
  1243

\bibitem[{{Fossum} \& {Carlsson}(2005)}]{fossum_05}
{Fossum}, A. \& {Carlsson}, M. 2005, \nat, 435, 919

\bibitem[{{Fossum} \& {Carlsson}(2006)}]{fossum_06}
{Fossum}, A. \& {Carlsson}, M. 2006, \apj, 646, 579

\bibitem[{{Hasan} \& {van Ballegooijen}(2008)}]{hasan_08}
{Hasan}, S.~S. \& {van Ballegooijen}, A.~A. 2008, \apj, 680, 1542

\bibitem[{{Holmes} {et~al.}(1996){Holmes}, {Lumley}, \& {Berkooz}}]{lum_96}
{Holmes}, P., {Lumley}, J.~M., \& {Berkooz}, G. 1996, {Turbulence, Coherent
  Structures, Dynamical Systems and Symmetry} ({Cambridge, England}: {Cambridge
  University Press})

\bibitem[{{Janssen} \& {Cauzzi}(2006)}]{janssen_06}
{Janssen}, K. \& {Cauzzi}, G. 2006, \aap, 450, 365

\bibitem[{{Jefferies} {et~al.}(2006){Jefferies}, {McIntosh}, {Armstrong},
  {Bogdan}, {Cacciani}, \& {Fleck}}]{jefferies_06}
{Jefferies}, S.~M., {McIntosh}, S.~W., {Armstrong}, J.~D., {et~al.} 2006,
  \apjl, 648, L151

\bibitem[{{Judge} {et~al.}(2003){Judge}, {Carlsson}, \& {Stein}}]{judge_03}
{Judge}, P.~G., {Carlsson}, M., \& {Stein}, R.~F. 2003, \apj, 597, 1158

\bibitem[{{Judge} \& {Carpenter}(1998)}]{judge_98}
{Judge}, P.~G. \& {Carpenter}, K.~G. 1998, \apj, 494, 828

\bibitem[{{Judge} {et~al.}(2004){Judge}, {Saar}, {Carlsson}, \&
  {Ayres}}]{judge_04}
{Judge}, P.~G., {Saar}, S.~H., {Carlsson}, M., \& {Ayres}, T.~R. 2004, \apj,
  609, 392

\bibitem[{{Judge} {et~al.}(2001){Judge}, {Tarbell}, \& {Wilhelm}}]{judge_01}
{Judge}, P.~G., {Tarbell}, T.~D., \& {Wilhelm}, K. 2001, \apj, 554, 424

\bibitem[{{Kalkofen}(2001)}]{kalkofen_01}
{Kalkofen}, W. 2001, \apj, 557, 376

\bibitem[{{Kalkofen} {et~al.}(1999){Kalkofen}, {Ulmschneider}, \&
  {Avrett}}]{kalkofen_99}
{Kalkofen}, W., {Ulmschneider}, P., \& {Avrett}, E.~H. 1999, \apjl, 521, L141

\bibitem[{{Kamio} \& {Kurokawa}(2006)}]{kamio_03}
{Kamio}, S. \& {Kurokawa}, H. 2006, \aap, 450, 351

\bibitem[{{Krijger} {et~al.}(2001){Krijger}, {Rutten}, {Lites}, {Straus},
  {Shine}, \& {Tarbell}}]{kri_01}
{Krijger}, J.~M., {Rutten}, R.~J., {Lites}, B.~W., {et~al.} 2001, \aap, 379,
  1052

\bibitem[{{Lawrence} {et~al.}(2004){Lawrence}, {Cadavid}, \&
  {Ruzmaikin}}]{ruzm_04}
{Lawrence}, J.~K., {Cadavid}, A., \& {Ruzmaikin}, A. 2004, \solphys, 225, 1

\bibitem[{{Leenaarts} {et~al.}(2006){Leenaarts}, {Rutten}, {S{\"u}tterlin},
  {Carlsson}, \& {Uitenbroek}}]{leenaarts_06}
{Leenaarts}, J., {Rutten}, R.~J., {S{\"u}tterlin}, P., {Carlsson}, M., \&
  {Uitenbroek}, H. 2006, \aap, 449, 1209

\bibitem[{{Lites} {et~al.}(1999){Lites}, {Rutten}, \& {Berger}}]{lites_99}
{Lites}, B.~W., {Rutten}, R.~J., \& {Berger}, T.~E. 1999, \apj, 517, 1013

\bibitem[{{Lites} {et~al.}(1994){Lites}, {Rutten}, \& {Thomas}}]{lites_94}
{Lites}, B.~W., {Rutten}, R.~J., \& {Thomas}, J.~H. 1994, in Solar Surface
  Magnetism, ed. R.~J. {Rutten} \& C.~J. {Schrijver}, 159--+

\bibitem[{{Liu}(1974)}]{liu_74}
{Liu}, S.-Y. 1974, \apj, 189, 359

\bibitem[{{Mart{\'{\i}}nez-Sykora} {et~al.}(2008){Mart{\'{\i}}nez-Sykora},
  {Hansteen}, \& {Carlsson}}]{martinez_07}
{Mart{\'{\i}}nez-Sykora}, J., {Hansteen}, V., \& {Carlsson}, M. 2008, \apj,
  679, 871

\bibitem[{{McIntosh} {et~al.}(2003){McIntosh}, {Fleck}, \&
  {Judge}}]{mcintosh_03}
{McIntosh}, S.~W., {Fleck}, B., \& {Judge}, P.~G. 2003, \aap, 405, 769

\bibitem[{{McIntosh} \& {Judge}(2001)}]{mcintoshjud_01}
{McIntosh}, S.~W. \& {Judge}, P.~G. 2001, \apj, 561, 420

\bibitem[{{Narain} \& {Ulmschneider}(1996)}]{narain_96}
{Narain}, U. \& {Ulmschneider}, P. 1996, Space Science Reviews, 75, 453

\bibitem[{{Neckel}(1999)}]{neckel_99}
{Neckel}, H. 1999, \solphys, 184, 421

\bibitem[{{Noyes}(1967)}]{noyes_67}
{Noyes}, R.~W. 1967, in IAU Symposium, Vol.~28, Aerodynamic Phenomena in
  Stellar Atmospheres, ed. R.~N. {Thomas}, 293--+

\bibitem[{{Orrall}(1966)}]{orrall_66}
{Orrall}, F.~Q. 1966, \apj, 143, 917

\bibitem[{{Pietarila} {et~al.}(2006){Pietarila}, {Socas-Navarro}, {Bogdan},
  {Carlsson}, \& {Stein}}]{pietarila_06}
{Pietarila}, A., {Socas-Navarro}, H., {Bogdan}, T., {Carlsson}, M., \& {Stein},
  R.~F. 2006, \apj, 640, 1142

\bibitem[{{Reardon} \& {Cavallini}(2008)}]{reardon_cavallini_08}
{Reardon}, K.~P. \& {Cavallini}, F. 2008, \aap, 481, 897

\bibitem[{{Reardon} {et~al.}(2008){Reardon}, {Lepreti}, {Carbone}, \&
  {Vecchio}}]{reardon_turbulence_08}
{Reardon}, K.~P., {Lepreti}, F., {Carbone}, V., \& {Vecchio}, A. 2008, \apjl,
  683, L207

\bibitem[{{Rees} {et~al.}(2000){Rees}, {L{\'o}pez Ariste}, {Thatcher}, \&
  {Semel}}]{rees_00}
{Rees}, D.~E., {L{\'o}pez Ariste}, A., {Thatcher}, J., \& {Semel}, M. 2000,
  \aap, 355, 759

\bibitem[{{Rezaei} {et~al.}(2008){Rezaei}, {Bruls}, {Schmidt}, {Beck},
  {Kalkofen}, \& {Schlichenmaier}}]{rezaei_08}
{Rezaei}, R., {Bruls}, J.~H.~M.~J., {Schmidt}, W., {et~al.} 2008, \aap, 484,
  503

\bibitem[{{Rezaei} {et~al.}(2007){Rezaei}, {Schlichenmaier}, {Beck}, {Bruls},
  \& {Schmidt}}]{rezaei_07}
{Rezaei}, R., {Schlichenmaier}, R., {Beck}, C.~A.~R., {Bruls}, J.~H.~M.~J., \&
  {Schmidt}, W. 2007, \aap, 466, 1131

\bibitem[{{Rimmele}(2004)}]{rimmele_04}
{Rimmele}, T.~R. 2004, in Advancements in Adaptive Optics. Proceedings of the
  SPIE, Volume 5490, ed. D.~{Bonaccini Calia}, B.~L. {Ellerbroek}, \&
  R.~{Ragazzoni}, 34--46

\bibitem[{{Rutten}(2007)}]{rutten_07}
{Rutten}, R.~J. 2007, in Astronomical Society of the Pacific Conference Series,
  Vol. 368, The Physics of Chromospheric Plasmas, ed. P.~{Heinzel},
  I.~{Dorotovi{\v c}}, \& R.~J. {Rutten}, 27--+

\bibitem[{{Rutten} \& {Uitenbroek}(1991)}]{rutten_91}
{Rutten}, R.~J. \& {Uitenbroek}, H. 1991, \solphys, 134, 15

\bibitem[{{Scherrer} {et~al.}(1995){Scherrer}, {Bogart}, {Bush}, {Hoeksema},
  {Kosovichev}, {Schou}, {Rosenberg}, {Springer}, {Tarbell}, {Title},
  {Wolfson}, {Zayer}, \& {MDI Engineering Team}}]{scherrer_95}
{Scherrer}, P.~H., {Bogart}, R.~S., {Bush}, R.~I., {et~al.} 1995, \solphys,
  162, 129

\bibitem[{{Schwarzschild}(1948)}]{schwarzschild_48}
{Schwarzschild}, M. 1948, \apj, 107, 1

\bibitem[{{Sivaraman} {et~al.}(2000){Sivaraman}, {Gupta}, {Livingston},
  {Dam{\'e}}, {Kalkofen}, {Keller}, {Smartt}, \& {Hasan}}]{sivaraman_00}
{Sivaraman}, K.~R., {Gupta}, S.~S., {Livingston}, W.~C., {et~al.} 2000, \aap,
  363, 279

\bibitem[{{Skartlien} {et~al.}(1994){Skartlien}, {Carlsson}, \&
  {Stein}}]{skartlien_94}
{Skartlien}, R., {Carlsson}, M., \& {Stein}, R.~F. 1994, in Chromospheric
  Dynamics, ed. M.~{Carlsson}, 79

\bibitem[{{Steffens} {et~al.}(1996){Steffens}, {Hofmann}, \&
  {Deubner}}]{steffens_96}
{Steffens}, S., {Hofmann}, J., \& {Deubner}, F.~L. 1996, \aap, 307, 288

\bibitem[{{Straus} {et~al.}(2008){Straus}, {Fleck}, {Jefferies}, {Cauzzi},
  {McIntosh}, {Reardon}, {Severino}, \& {Steffen}}]{straus_08}
{Straus}, T., {Fleck}, B., {Jefferies}, S.~M., {et~al.} 2008, \apjl, 681, L125

\bibitem[{{Tritschler} {et~al.}(2007){Tritschler}, {Schmidt}, {Uitenbroek}, \&
  {Wedemeyer-B{\"o}hm}}]{tritschler_07}
{Tritschler}, A., {Schmidt}, W., {Uitenbroek}, H., \& {Wedemeyer-B{\"o}hm}, S.
  2007, \aap, 462, 303

\bibitem[{{Ulmschneider} \& {Musielak}(2003)}]{ulmrew_03}
{Ulmschneider}, P. \& {Musielak}, Z. 2003, in Astronomical Society of the
  Pacific Conference Series, Vol. 286, Current Theoretical Models and Future
  High Resolution Solar Observations: Preparing for ATST, ed. A.~A. {Pevtsov}
  \& H.~{Uitenbroek}, 363--+

\bibitem[{{Ulmschneider} {et~al.}(2005){Ulmschneider}, {Rammacher}, {Musielak},
  \& {Kalkofen}}]{ulmschneider_05}
{Ulmschneider}, P., {Rammacher}, W., {Musielak}, Z.~E., \& {Kalkofen}, W. 2005,
  \apjl, 631, L155

\bibitem[{{Vecchio}(2006)}]{vecchio2}
{Vecchio}, A. 2006, \aap, 446, 669

\bibitem[{{Vecchio} {et~al.}(2005){Vecchio}, {Carbone}, {Lepreti}, {Primavera},
  {Sorriso-Valvo}, {Veltri}, {Alfonsi}, \& {Straus}}]{vecchio1}
{Vecchio}, A., {Carbone}, V., {Lepreti}, F., {et~al.} 2005, Physical Review
  Letters, 95, 061102

\bibitem[{{Vecchio} {et~al.}(2007){Vecchio}, {Cauzzi}, {Reardon}, {Janssen}, \&
  {Rimmele}}]{noi_07}
{Vecchio}, A., {Cauzzi}, G., {Reardon}, K.~P., {Janssen}, K., \& {Rimmele}, T.
  2007, \aap, 461, L1

\bibitem[{{Vernazza} {et~al.}(1981){Vernazza}, {Avrett}, \&
  {Loeser}}]{avrett81}
{Vernazza}, J.~E., {Avrett}, E.~H., \& {Loeser}, R. 1981, \apjs, 45, 635

\bibitem[{{von Uexkuell} \& {Kneer}(1995)}]{vonuexkull_95}
{von Uexkuell}, M. \& {Kneer}, F. 1995, \aap, 294, 252

\bibitem[{{Wedemeyer} {et~al.}(2004){Wedemeyer}, {Freytag}, {Steffen},
  {Ludwig}, \& {Holweger}}]{wedemeyer_04}
{Wedemeyer}, S., {Freytag}, B., {Steffen}, M., {Ludwig}, H.-G., \& {Holweger},
  H. 2004, \aap, 414, 1121

\bibitem[{{Wedemeyer-B{\"o}hm} {et~al.}(2007){Wedemeyer-B{\"o}hm}, {Steiner},
  {Bruls}, \& {Rammacher}}]{wedemeyeretal_07}
{Wedemeyer-B{\"o}hm}, S., {Steiner}, O., {Bruls}, J., \& {Rammacher}, W. 2007,
  in Astronomical Society of the Pacific Conference Series, Vol. 368, The
  Physics of Chromospheric Plasmas, ed. P.~{Heinzel}, I.~{Dorotovi{\v c}}, \&
  R.~J. {Rutten}, 93--+

\bibitem[{{Wedemeyer-B{\"o}hm} \& {W{\"o}ger}(2008)}]{wedemeyer_07}
{Wedemeyer-B{\"o}hm}, S. \& {W{\"o}ger}, F. 2008, in IAU Symposium, Vol. 247,
  IAU Symposium, 66--73

\bibitem[{{Wikst{\o}l} {et~al.}(2000){Wikst{\o}l}, {Hansteen}, {Carlsson}, \&
  {Judge}}]{wikstol_00}
{Wikst{\o}l}, {\O}., {Hansteen}, V.~H., {Carlsson}, M., \& {Judge}, P.~G. 2000,
  \apj, 531, 1150

\bibitem[{{Wilhelm} {et~al.}(1995){Wilhelm}, {Curdt}, {Marsch}, {Sch{\"u}hle},
  {Lemaire}, {Gabriel}, {Vial}, {Grewing}, {Huber}, {Jordan}, {Poland},
  {Thomas}, {K{\"u}hne}, {Timothy}, {Hassler}, \& {Siegmund}}]{wilhelm_95}
{Wilhelm}, K., {Curdt}, W., {Marsch}, E., {et~al.} 1995, \solphys, 162, 189

\bibitem[{{W{\"o}ger}(2007)}]{woeger_06}
{W{\"o}ger}, F. 2007, PhD thesis, Kiepenheuer-Institut f{\"u}r Sonnenphysik
  Albert-Ludwigs-University, Freiburg, Germany

\end{thebibliography}
\end{document}